\documentclass{aastex}          
\usepackage{spr-astr-addons}    
\usepackage{rotating}

\newcommand{\wse}{{\it WISE}}

\newcommand{\fer}{{\it Fermi}}

\begin{document}
%
\title{Optical Spectroscopic Observations of Gamma-Ray Blazar Candidates VIII: The 2016-2017 follow up campaign carried out at SPM, NOT, KPNO and SOAR telescopes.}

\shorttitle{Optical Spectroscopic Observations of Gamma-Ray Blazar Candidates VIII}
\shortauthors{Marchesini, E. J. et al.}
\email{ezmarche@unito.it}

\author{Marchesini, E. J. \altaffilmark{1,2,3,4,5}} \and \author{Pe\~na-Herazo, H. A. \altaffilmark{3,4,6}} \and \author{\'Alvarez Crespo, N. \altaffilmark{3,4}} \and \author{Ricci, F.\altaffilmark{7,8,9}} \and \author{Negro, M. \altaffilmark{3,4,10}} \and \author{Milisavljevic, D. \altaffilmark{11}} \and \author{Massaro, F. \altaffilmark{3,4,10}} \and \author{Masetti, N.\altaffilmark{5,12}} \and \author{Landoni, M.\altaffilmark{13}} \and \author{Chavushyan, V.\altaffilmark{6}} \and \author{D'Abrusco, R.\altaffilmark{7}} \and \author{Jim\'enez-Bail\'on, E.\altaffilmark{14}} \and \author{La Franca, F.\altaffilmark{8}} \and \author{Paggi, A.\altaffilmark{3,4,10}} \and \author{Smith, H. A.\altaffilmark{7}} \and \author{Tosti, G.\altaffilmark{15}}


\altaffiltext{1}{Facultad de Ciencias Astron\'omicas y Geof\'isicas, Universidad Nacional de La Plata, La Plata, Argentina.}
\altaffiltext{2}{Instituto de Astrof\'isica de La Plata, CONICET-UNLP, CCT La Plata, La Plata, Argentina.}
\altaffiltext{3}{Dipartimento di Fisica, Universita degli Studi di Torino, Torino, Italy.}
\altaffiltext{4}{INFN - Istituto Nazionale di Fisica Nucleare, Sezione di Torino, Torino, Italy.}
\altaffiltext{5}{INAF - Osservatorio di Astrofisica e Scienza dello Spazio di Bologna, Bologna, Italy.}
\altaffiltext{6}{Instituto Nacional de Astrof\'isica, \'Optica y Electr\'onica, Puebla, Mexico.}
\altaffiltext{7}{Harvard - Smithsonian Center for Astrophysics, Cambridge, MA, USA.}
\altaffiltext{8}{Dipartimento di Matematica e Fisica, Universit\`a Roma Tre, Roma, Italy.}
\altaffiltext{9}{Instituto de Astrof\'isica and Centro de Astroingenier\'ia, Facultad de F\'isica, Pontificia Universidad Cat\'olica de Chile, Casilla 306, Santiago 22, Chile.}
\altaffiltext{10}{INAF - Osservatorio Astrofisico di Torino, Pino Torinese, Italy.}
\altaffiltext{11}{Department of Physics and Astronomy, Purdue University, West Lafayette, IN, USA.}
\altaffiltext{12}{Departamento de Ciencias F\'isicas, Universidad Andr\'es Bello, Santiago, Chile.}
\altaffiltext{13}{INAF - Osservatorio Astronomico di Brera, Merate, Italy.}
\altaffiltext{14}{Instituto de Astronom\'ia, Universidad Nacional Aut\'onoma de M\'exico, Baja California, Mexico.}
\altaffiltext{15}{Dipartimento di Fisica, Universit\`a degli Studi di Perugia, Perugia, Italy.}

\begin{abstract}
The third \fer \,\,source catalog lists 3033 $\gamma$-ray sources above $4\sigma$ significance. More than 30\% are classified as either unidentified/unassociated Gamma-ray sources (UGSs), with about 20\% classified as Blazar candidates of uncertain types (BCUs). To confirm the blazar-like nature of candidate counterparts of UGSs and BCUs, we started in 2012 an optical spectroscopic follow up campaign. We report here the spectra of 36 targets with observations from the Observatorio Astron\'omico Nacional San Pedro M\'artir, the Southern Astrophysical Research Observatory, the Kitt Peak National Observatory and the Northern Optical Telescope, between 2016 and 2017.
We confirm the BL Lac nature of 23 sources, and the flat spectrum radio quasar nature of other 7 ones. We also provide redshift estimates for 19 out of these 30 confirmations, with only one being a lower limit due to spectral features ascribable to intervening systems along the line of sight. As in previous analyses, the largest fraction of now-classified BCUs belong to the class of BL Lac objects, that appear to be the most elusive class of active galactic nuclei. One of the BL Lacs identified in this work, associated with 3FGL J2213.6-4755, lies at a redshift of $z>$1.529, making it one of the few distant gamma-ray BL Lac objects.
\end{abstract}


\section{Introduction}

We are currently living in a golden age for gamma-ray astronomy. Thanks to the discoveries performed with the Large Area Telescope (LAT) \citep[see e.g.,][for a recent review on the extragalactic gamma-ray sky]{review}, on board the \fer\ Gamma-ray Space Telescope \citep{fermilat} launched in 2008, the number of gamma-ray sources is continuously increasing. There are 1450 gamma-ray objects listed in the \fer-LAT First Source Catalog \citep[1FGL;][]{1fgl} and more than 3000 included in the last release of the \fer-LAT Third Source Catalog \citep[3FGL;][]{3fgl}.

However, one of the key scientific questions identified before the launch of \fer\ is still unsolved: a significant fraction of the objects detected by \fer\, ranging between 30\% to 40\%, still constitute the population of the unidentified/unassociated gamma-ray sources (UGSs) lacking a classified, low-energy counterpart \citep[see e.g.,][]{egret,agile}. The search for UGS counterparts at lower frequencies is a challenging task \citep[see e.g.,][and references therein]{opt7}, mostly due to the \emph{Fermi}-LAT positional uncertainty that is at least an order of magnitude larger than that in the radio-infrared-optical and even soft X-ray band \citep{ugshunt,ugs5kde}. This prevents us from using standard procedures for positional crossmatches, making it necessary to use multifrequency information \citep[see e.g,][]{archival,ugs2mulfrq,blarch} and associated statistical analyses.

Blazars are among the most extreme classes of radio-loud active galactic nuclei (AGN) whose observational properties and emission at all wavelengths are generally interpreted as due to particles accelerated in a relativistic jet aligned, within a small angle, to our line of sight \citep{blandford}. Blazars emit via non-thermal processes over the entire electromagnetic spectrum, from radio to $\gamma$ rays \citep[e.g.,][]{urrypadovani}. They exhibit a broad, double-bump spectral energy distribution (SED) coupled with highly variable (i.e., days to minute time scales) and polarized emission \citep{andru05,angelakis16}, apparent superluminal motions \citep{jorstad01}, flat radio spectra, even at low radio frequencies \citep[see e.g.][]{bl74} and peculiar infrared colors \citep{strip,irbl,gir} discovered thanks to the all-sky survey carried out with the \wse\ satellite \citep{wright10}.

Blazars are mainly classified as BL Lac objects, from the name of the prototype, when the equivalent width (EqW) of the emission and/or absorption lines present in their optical spectra is less than 5\,\AA\ \citep{stickel91}, while they are known as flat spectrum radio quasars when broad emission lines typical of quasar-like (QSO) spectra are visible. The faintness of spectral features, if present at all, in BL Lac objects is the reason why the determination of their redshifts is challenging \citep{landoni15,Paiano17b,Paiano17c}.

There are several methods developed to properly assess the blazar nature of a $\gamma$-ray source and/or to search for its potential low-energy counterpart, as those based on radio \citep{ghirlanda10,mwabl}, infrared \citep{ugs1met}, optical \citep{sandrinelli13,marchesini16,Paiano17d}, optical polarization \citep{robopol}, X-ray follow up observations \citep{stephen10,Takahashi12,Takeuchi13,ugs4xray,landi15}, a broader multiwavelength approach \citep{Paiano17a}, or machine learning algorithms \citep{Doert14}. Unfortunately to get a conclusive answer on the nature of the low-energy counterpart associated with a \fer\ source, optical spectroscopic observations are strictly necessary \citep{refined}. This is also the case of those sources listed in the \fer\ AGN catalogs \citep[1LAC, 2LAC and 3LAC][]{1lac,2lac,3lac} known as blazar candidate of uncertain type (BCUs).

BCUs are radio, infrared or X-ray sources associated with a \fer\ source on the basis of the statistical procedures developed in the \fer\ catalogs \citep{3lac}. If there is a positional correlation between a gamma-ray souce and a lower energy candidate counterpart, which is in turn either an unclassified blazar, or an unidentified source with a typical two-humped blazar-like spectral energy distribution, then this candidate counterpart is classified as a BCU. In particular, this means that these potential counterparts of $\gamma$-ray sources show multifrequency behavior similar to blazars, satisfying at least one of the following criteria: (i) are listed as blazars of uncertain type (BZUs) in the Roma-BZCAT \citep{bzcat09,bzcat15} or (ii) are sources with a flat radio spectrum and/or showing the typical two-humped SED and listed in at least one of the catalogs used to associate $\gamma$-ray sources in the \fer\ catalogs.

In 2012, we started an optical spectroscopic campaign to perform follow up observations of UGSs for which our statistical algorithms based on the infrared colors \citep{agus,ugs1met} identified a potential counterparts \citep[see e.g.,][for the initial results of our campaign]{opt1,sdss}, as well as to search for a blazar confirmation on the nature of the BCUs \citep{optbcu}. To date more than 250 new blazars have been discovered thanks to our optical campaign \citep[see e.g.,][for a recent review]{quest} and will be listed in the future releases of the \fer\ gamma-ray catalogs.

Results of our follow up optical campaign were used by several groups to: (i) build the luminosity function of BL Lacs \citep[see e.g.,][]{ajello14}; (ii) select potential targets for the Cherenkov Telescope Array \citep[see e.g.,][]{tevcan,arsioli15}; (iii) obtain stringent limits on the dark matter annihilation in sub-halos \citep[see e.g.,][]{zechlin12,berlin14}; (iv) search for counterparts of new flaring gamma-ray sources \citep[see e.g.,][]{bernieri13}; (v) test new gamma-ray detection algorithms \citep[see e.g.,][]{campana15,campana16,campana17}; (vi) perform population studies on the UGSs \citep[see e.g.,][]{acero13} and (vii) discover the new subclass of radio weak BL Lacs \citep[see e.g.,][]{rwbls}, to name a few.

Here we present an analysis of the spectroscopic observations collected between 2016 and 2017 on a sample of BCUs. The current paper is organized as follows. In \S~{2.1} we present the sample selection, then in \S~{2.2} we briefly describe the data reduction procedures. \S~{3} is devoted to our results while summary and conclusions are given in \S~{4}. Unless otherwise stated we adopt cgs units for numerical results and spectral indices, $\alpha$, are defined by flux density, S$_{\nu}\propto\nu^{-\alpha}$. We also adopt a cosmological model with $H_0=70$, $\Omega_M=0.3$ and $\Omega_{\Lambda}=0.7$.

\section{Observations}

\subsection{Sample selection}
 
To be consistent with our previous analyses, throughout this paper we will adopt the Roma-BZCAT \citep{bzcat09,bzcat15} nomenclature: labeling as BZB the BL Lac subclass, and as BZQ the Flat spectrum radio quasars subclass.

As in previous analyses, we built a list of possible targets out of the potential counterparts of the UGSs with IR colors \citep[see also][]{wibrals1} or low radio frequency spectra \citep[below $\sim$ 1 GHz][]{ugs3lowfrq,ugs6lowfrq,mwabl} similar to known, associated, \fer\ blazars; or out of the associated candidate counterparts to BCUs as listed by the Fermi catalogs \citep{3lac,3fgl}. Out of those targets, we selected those that were visible during the observing nights awarded to our optical spectroscopic campaign. The final sample is further reduced after taking into account magnitude limits for each observatory, which in our case range from 18 to 19.5 magnitudes in the R band.

We observed 30 targets associated with \fer\ gamma-ray sources and classified as BCUs in the 3LAC \citep{3lac}, then we also observed 4 BZBs already present in the latest version of the Roma-BZCAT for which a spectroscopic redshift estimate is not reported in the literature. In the latter case given the variabile emission of BL Lacs we re-observed these sources hoping to find them in a low/quiescent state so to potentially detect some emission and/or absorption features in their optical spectra, the same strategy we adopted in the past to fill gaps in our schedule during the observations \citep{opt2,opt3}.

We also observed one BL Lac candidate, as classified in the Roma-BZCAT, that is a source classified as BL Lac but not confirmed, i.e., with only a description of its optical spectrum available in the literature but no images to verify it. This source, 5BZB J0607+4739, has been also associated with the \fer\ object: 3FGL\,J0607.4+4739 \citep{refined}. Finally, we also observed one source listed as a potential blazar-like counterpart of the UGS 3FGL\,J0952.8+0711, a blazar candidate indicated by our procedures based on the IR colors of gamma-ray blazars \citep{ugs1met}. In summary, our sample includes a total of 36 sources. 

It is worth noticing that, given that they are BCUs, some information was already known beforehand. In some cases, the source was already classified as an AGN due to its physical properties but a proper classification (as either BL Lac or FSRQ) was missing. For example, in the cases of 3FGL J1816.9-4944 and 3FGL J1911.4-1908 \citep[and references therein]{refined}, for which an optical spectral classification was necessary to confirm their nature. In the case of 3FGL J0653.6+2817, optical data was already available in the literature but was non-conclusive \citep{Mahony11,opt5}. The BZB sources are the only ones already classified, for which we aim only to pinpoint their redshifts. Finally, a redshift of 0.591 was already reported in the literature for the candidate counterpart of object 3FGL J0922.8-3959 \citep{White88}, although there were no optical spectra available. From our spectrum, the object shows strong emission lines, at a redshift of $z=0.595$, which is slightly different but still compatible with the previous value. Lastly, an optical spectroscopic identification for the UGS source 3FGL\,J0952.8+0711 was later published in the literature \citep{Paiano17d}. We nevertheless checked whether this source showed any signs of spectral variability. As for the rest of our sample, no optical information whatsoever was found.

\subsection{Data reduction}

All our data were acquired, reduced and analyzed following the same method adopted in our previous analyses of this spectroscopic campaign. Here we reported only the basic details regarding the data collection and a brief overview of our standard procedure used for the extraction of the optical spectra, more information can be found in our previous works \citep[see e.g.,][for additional details]{opt4,opt5,opt6,opt7}.

Targets were observed at several observatories:
\begin{itemize}
\item Three sources were observed with the 2.1\,m telescope at the San Pedro Martir (SPM) Observatory, Mexico, on two runs: on 2016/01/16 the first one and between 2017/08/22 and 2017/08/24 the second one. We used the Boller \& Chivens spectrograph with a resolution of 2.3\,\AA\ and a spectral range of 3800\,\AA\ to 8200\,\AA.
\item Twenty-four sources were observed with the 4.1\,m telescope from the Southern Astrophysical Research Observatory (SOAR), in Chile, on two runs between 2017/05/08 and 2017/05/09, and between 2017/08/18 and 2017/08/20, respectively. We used the Goodman spectrograph with a resolution of 2\,\AA\ and a spectral range from 4000\,\AA\ to 8000\AA.
\item One source was observed with the 2.5\,m Northern Optical Telescope (NOT), in 2016/01/14. We used the ALFOSC spectrograph with a resolution of 7\,\AA\ and a spectral range covering from 4200\,\AA\ to 9700\,\AA.
\item Eight sources were observed with the 4\,m telescope at the Kitt Peak National Observatory (KPNO), USA, between 2016/08/12 and 2016/10/12. We used the KOSMOS spectrograph with a resolution of 1\,\AA\ and a spectral range covering from 5000\,\AA\ to 8200\,\AA.
\end{itemize}

All the data were reduced using the standard IRAF procedures for bias and flat correction, for cosmic ray and sky subtraction, and for wavelength and flux calibration. Wavelength calibration was performed with spectra from comparison lamps taken immediately after or right before the pointing of the object of interest, to minimize flexion issues, if any. Spectrophotometric standards were taken every night with the same configuration, for flux calibration purposes. To remove cosmic rays we used the L.A.Cosmic IRAF algorithm \citep{lacos}, which provides a robust identification and removal of cosmic rays. 
In the cases in which more than one exposure was taken (to ensure minimum influence of the cosmic rays over the final result), the final spectra were combined. All the final spectra were dereddened for galactic absorption assuming $E_{(B-V)}$ values as given by the NASA/IPAC Galactic Dust\footnote{http://irsa.ipac.caltech.edu/applications/DUST/}, Reddening and Extinction maps \citep{dered}, and normalized to better visualize spectral features (if any).

All the details about our sample together with the log of our observations are presented in Table 1, where we report the \fer\ name and that of its assigned counterpart taken from the 3LAC catalog \citep{3lac}, the \wse\ name, its classification based on the spectra collected, an estimated value for its redshift (for those sources showing emission/absorption lines), the observing dates, the average signal-to-noise ratio (SNR) and the exposure time. The spectra presented in this work show all SNRs of at least 10, to strengthen the confidence in our results. Spectra with an SNR of less than 30 are flagged, to indicate that the resulting classification should be taken with caution. We claim a line identification whenever there are at least two absorption and/or emission lines present, that can be identified with a single redshift value. The only exceptions to this are 3FGL J2107.7-4822 and 3FGL J2213.6-4755, for which the resulting redshift are only a lower limit.

\begin{sidewaystable*}
\label{tab:main}
\small

\begin{tabular}{|l|l|l|c|c|l|l|l|l|l|}
\hline
  \multicolumn{1}{|c|}{Fermi Name} &
  \multicolumn{1}{c|}{Association} &
  \multicolumn{1}{c|}{WISE Name} &
  \multicolumn{1}{c|}{Fermi} &
  \multicolumn{1}{c|}{BZCat} &
  \multicolumn{1}{c|}{Redshift} &
  \multicolumn{1}{c|}{Obs. date} &
  \multicolumn{1}{c|}{SNR} &
  \multicolumn{1}{c|}{Exp. time} &
  \multicolumn{1}{c|}{Obs.} \\
  & & & class & class & & [yyyy-mm-dd] &  & [s] & \\
\hline
  3FGL J0021.6-6835$^*$ & PKS 0021-686 & J002406.72-682054.5 & BCU & QSO & 0.354 & 2017-08-20 & 11 & 2x600 & SOAR \\
  3FGL J0040.5-2339$^*$ & PMN J0040-2340 & J004024.90-234000.7 & BCU & BZB/galaxy & 0.213 & 2017-08-18 & 14 & 2x700 & SOAR \\
  3FGL J0050.0-4458 & PMN J0049-4457 & J004916.62-445711.2 & BCU & QSO/Sy & 0.121 & 2017-08-20 & 31 & 2x500 & SOAR \\
  3FGL J0059.1-5701$^*$ & PKS 0056-572 & J005846.59-565911.4 & BCU & QSO & 0.677 & 2017-08-18 & 15 & 2x1000 & SOAR\\
  3FGL J0133.2-5159$^*$ & PKS 0131-522 & J013305.75-520003.9 & BCU & QSO & 0.925 & 2017-08-18 & 17 & 2x1200 & SOAR \\
  3FGL J0211.2-0649$^*$ & NVSS J021116-064422 & J021116.95-064419.9 & BCU & BZB & 0.194 & 2017-08-20 & 17 & 2x500 & SOAR \\
  3FGL J0301.4-1652$^*$ & PMN J0301-1652 & J030116.62-165245.0 & BCU & BZB/galaxy & 0.278 & 2017-08-20 & 27 & 2x800 & SOAR \\
  3FGL J0623.3+3043 & B6 J0623+3045 & J062316.03+304457.7 & BCU & BZB & ? & 2016-12-09 & 30 & 2x1200.0 & KPNO \\
  3FGL J0653.6+2817$^*$ & B6 J0653+2816 & J065344.26+281547.5 & BCU & BZB & ? & 2016-12-10 & 25 & 2x900 & KPNO \\
  3FGL J0723.7+2050$^*$ & GB6 J0723+2051 & J072348.34+205130.8 & BCU & BZB & ? & 2016-01-14 & 22 & 3x600 & NOT \\
  3FGL J0748.8+4929 & VSS J074837+493040 & J074837.76+493041.0 & BCU & BZB & ? & 2016-12-09 & 75 & 2x1800 & KPNO \\
  3FGL J0922.8-3959$^*$ & PKS 0920-39 & J092246.42-395935.0 & BCU & QSO & 0.595 & 2017-05-08 & 11 & 3x600 & SOAR \\
  3FGL J1052.8-3741$^*$ & PMN J1053-3743 & J105258.09-374318.6 & BCU & BZB & ? & 2017-05-09 & 16 & 3x400 & SOAR \\
  3FGL J1440.0-3955$^*$ & 1RXS J143949.8-395524 & J143950.86-395518.8 & BCU & BZB & 0.300 & 2017-08-19 & 11 & 1x1000 & SOAR \\
  3FGL J1559.8-2525$^*$ & NVSS J160005-252439 & J160005.35-252439.7 & BCU & BZB & 0.416 & 2017-08-18 & 24 & 1x1200 & SOAR \\
  3FGL J1816.9-4944$^*$ & PMN J1816-4943 & J181655.99-494344.7 & BCU & QSO & 1.70 & 2017-08-20 & 27 & 2x1200 & SOAR \\
  3FGL J1842.3-5841$^*$ & 1RXS J184230.6-584202 & J184229.83-584157.5 & BCU & BZB & 0.421 & 2017-05-09 & 28 & 3x770 & SOAR \\
  3FGL J1911.4-1908 & PMN J1911-1908 & J191129.74-190824.8 & BCU & BZB/galaxy & 0.138 & 2017-08-20 & 38 & 3x870 & SOAR \\
  3FGL J1954.9-5640 & 1RXS J195503.1-564031 & J195502.86-564028.8 & BCU & BZB & 0.221 & 2017-08-18 & 42 & 2x900 & SOAR \\
  3FGL J2046.7-1011 & PMN J2046-1010 & J204654.33-101040.2 & BCU & BZB & ? & 2017-08-18 & 30 & 2x1000 & SOAR \\
  3FGL J2103.9-6233$^*$ & PMN J2103-6232 & J210338.38-623225.8 & BCU & BZB & ? & 2017-08-19 & 13 & 2x800 & SOAR \\
  3FGL J2107.7-4822$^*$ & PMN J2107-4827 & J210744.48-482802.9 & BCU & QSO & $>$0.519 & 2017-08-18 & 19 & 2x1200 & SOAR \\
  3FGL J2126.5-3926$^*$ & PMN J2126-3921 & J212625.19-392122.2 & BCU & BZB & ? & 2017-08-20 & 15 & 4x1300 & SOAR \\
  3FGL J2144.2+3132$^*$ & MG3 J214415+3132 & J214415.22+313339.2 & BCU & BZB & ? & 2017-08-22 & 23 & 3x1800 & SPM \\
  3FGL J2159.2-2841$^*$ & NVSS J215910-284115 & J215910.92-284116.4 & BCU & BZB & 0.270 & 2017-08-20 & 27 & 2x900 & SOAR \\
  3FGL J2212.6+2801$^*$ & MG3 J221240+2759 & J221239.11+275938.4 & BCU & BZB & ? & 2017-08-24 & 13 & 3x1800 & SPM \\
  3FGL J2213.6-4755$^*$ & SUMSS J221330-475426 & J221330.35-475425.2 & BCU & BZB & $>$1.529 & 2017-08-18 & 28 & 2x1200 & SOAR \\
  3FGL J2305.3-4219$^*$ & SUMSS J230512-421859 & J230512.44-421857.2 & BCU & BZB & ? & 2017-08-20 & 13 & 3x1200 & SOAR \\
  3FGL J2316.8-5209 & SUMSS J231701-521003 & J231701.72-521001.4 & BCU & BZB & 0.646 & 2017-08-18 & 32 & 3x1200 & SOAR \\
  3FGL J2348.4-5100 & SUMSS J234852-510311 & J234853.10-510314.0 & BCU & BZB & 0.392 & 2017-08-20 & 31 & 2x1200 & SOAR \\
\hline
  3FGL J0115.8+2519 & 5BZB J0115+2519 & J011546.15+251953.4 & BZB & BZB & ? & 2016-12-09 & 50 & 2x900 & KPNO \\
  3FGL J0540.4+5823 & 5BZB J0540+5823 & J054030.01+582338.4 & BZB & BZB & ? & 2016-12-10 & 25 & 2x1200 & KPNO \\
  3FGL J0625.2+4440 & 5BZB J0625+4440 & J062518.26+444001.6 & BZB & BZB & ? & 2016-12-08 & 50 & 2x900 & KPNO\\
  3FGL J0708.9+2239 & 5BZB J0708+2241 & J070858.28+224135.4 & BZB & BZB & ? & 2016-01-16 & 29 & 1x1800 & SPM \\
\hline
  3FGL J0607.4+4739 & 5BZB J0607+4739 & J060723.25+473947.0 & BZB & BZB & ? & 2016-12-10 & 70 & 2x900 & KPNO \\
\hline
  3FGL J0952.8+0711 & VSS J095249+071330 & J095249.57+071330.1 & UGS & BZB & 0.574$\dagger$ & 2016-12-09 & 15 & 2x900 & KPNO \\
\hline
\end{tabular}
\caption{Results from our campaign. In column 1 we report the 3FGL name; in column 2 the associated counterpart; in column 3 the WISE counterpart; in column 4 the Fermi class; in column 5 the Roma-BZCAT class we assigned to it from our data; in column 6 its redshift (as measured from our spectra, if any); in column 7 the observation date; in column 8 the signal to noise ratio; in column 9 the total number and length, in seconds, of exposures taken; and finally in column 10 we report the observatory at which the data were taken. It is worth noticing that the precise (with an uncertainty of less than 1\farcs) equatorial celestial coordinates in J2000.0 are given in the WISE denomination. $^*$Spectra with SNR less than 30 are marked with an asterisk, to indicate that the resulting classification should be taken with caution. $\dagger$The redshift value reported for this object is available on the literature, but in our spectrum there are no detected lines.}
\end{sidewaystable*}

\section{Results and Source details}

As occurred in our previous BCU counterpart analyses \citep{optbcu}, we found that a large  fraction of them are classifiable as BL Lacs. In particular, as shown in Table 1, 23 out of 30 BCUs pointed are BL Lacs, 11 with a certain redshift estimate, and 11 with a classical featureless spectrum. The remaining one is SUMSS J221330-475426, associated with 3FGL\,J2213.6-4755 and with a lower limit on its $z$ of 1.529. This lower limit is due to the presence of some absorption lines from intervening systems along the line of sight. There are also 3 out of the 12 BZBs, namely: PMN J0040-2340, PMN J0301-1652 and PMN J1911-1908, associated with 3FGL J0040.5-2339, 3FGL J0301.4-1652 and 3FGL J1911.4-1908, respectively, that appear to have the optical spectrum partially dominated by the one of their host galaxy and thus could resemble the BZG definition reported in the latest version of the Roma-BZCAT \citep[see also][for more details]{bzcat15}. It is worth mentioning that SUMSS J221330-475426, the associated counterpart the BCU 3FGL J2213.6-4755, is a BL Lac source with a measurable redshift of $>$1.529 based on FeII and MgII absorption doublets. As these doublets are very narrow, we conclude they must be from an intervening system, thus making the z value a lower limit. This means it is one of the few distant BZBs lying within the end of the distance distribution for BL Lacs \citep{bzcat09}. According to the Roma-BZCAT, only 15 BL Lac objects are reported at redshifts higher than 1, with 12 of them being uncertain. This is why there are a number of campaigns to find such objects \citep{Landoni18}, which can significantly help population studies. Lastly, in the case of SUMSS J231701-521003, associated with 3FGL J2316.8-5209, we detect two absorption doublets. They are compatible with the MnII and FeII resonance lines, the first, and with the CaII doublet the second, both at a redshift of $z=0.646$. However, the doublet in the blue end of the spectrum could be also attributed to the MgII lines from an intervening system at $z=0.517$.

The remaining 7 BCUs observed are all classifiable as QSO with a redshift estimate, with the only exception for PMN J0049-4457 associated with the \fer\ source 3FGL J0050.0-4458, which shows an optical spectrum more similar to that of the nucleus of a Seyfert galaxy 1.8 or 1.9 \citep[see e.g.,][]{masetti10,rojas17}, but has a clear flat radio spectrum and a bolometric luminosity typical of a BZQ (of the order of $10^{46}$\textrm{erg/s}). Moreover, although PMN J2107-4827, associated with 3FGL\,2107.7-4822, shows a broad emission feature in its spectrum and is thus classified as a QSO, we can only give a lower limit to its redshift due to an intervening system (i.e., $z > 0.519$) from which we detect the MgII double absorption. The emission line, although certainly from the object itself, remains unidentified.

Our optical spectrum of 5BZB\,J0607+4739, the BL Lac candidate listed in the Roma-BZCAT and associated with 3FGL\,J0607.4+4739, is featureless thus suggesting it is classifiable as BL Lac; as occurs for all the other four BZBs, namely: 5BZB\,J0115+2519, 5BZB\,J0540+5823, 5BZB\,J0625+4440 and 5BZB\,J0708+2241, for which, unfortunately, it was not possible to obtain a $z$ estimate.

Finally, the low-energy source VSS\,J095249+071330 that lies within the positional uncertainty region of the UGS 3FGL\,J0952.8+0711 is a classical BZB with a featureless optical spectrum. This is in agreement with the spectrum already reported in the literature \cite{Paiano17d}, although in our case no lines are visible. This is probably due to the lower resolution of our spectrum. 

All the lines measured from the observed spectra are reported in Table 2 for each 3FGL counterpart, alongside their respective line identifications.

\begin{table*}
\begin{center}

\begin{tabular}{|l|c|c|c|r|l|}
\hline\hline
  \multicolumn{1}{|c|}{Name} &
  \multicolumn{1}{c|}{z} &
  \multicolumn{1}{c|}{Line} &
  \multicolumn{1}{c|}{Obs. wavelength} &
  \multicolumn{1}{c|}{Eqw.} &
  \multicolumn{1}{c|}{Eqw. error} \\
  
  \multicolumn{1}{|c|}{[3FGL]} &
  \multicolumn{1}{c|}{} &
  \multicolumn{1}{c|}{} &
  \multicolumn{1}{c|}{[\AA ]} &
  \multicolumn{1}{c|}{[\AA ]} &
  \multicolumn{1}{c|}{[\AA ]} \\
\hline\hline
  3FGL J0021.6-6835 & 0.354 & [NeVI] & 4640 & +2.1 & 0.1\\
   &  & [OII]d & 5040 & +6.9 & 0.2\\
   &  & [NeIII] & 5240 & +3.1 & 0.1\\
   &  & HeI & 5267 & +1.6 & 0.1\\
   &  & H$\epsilon$ & 5376 & +4.1 & 0.2\\
   &  & H$\delta$ & 5555 & +18 & 1\\
   &  & H$\gamma$ & 5882 & +29.8 & 0.9\\
   &  & FeII & $\sim$6160 & - & -\\
   &  & H$\beta$ & 6584 & +57 & 3\\
   &  & [OIII] & 6717 & +13.8 & 0.4\\
   &  & [OIII] & 6782 & +53 & 1\\
   &  & FeII & 7180 & - & \\
   \hline
  3FGL J0040.5-2339 & 0.213 & CaII (K) & 4772 & 5.0 & 0.2\\
   &  & CaII (H) & 4813 & 3.8 & 0.2\\
   &  & G band & 5220 & 2.8 & 0.2\\
   &  & MgI & 6280 & 3.3 & 0.2\\
   \hline
  3FGL J0050.0-4458 & 0.121 & [NeIII] & 4338 & +2.5 & 0.2\\
   &  & CaII (K) & 4411 & 3.5 & 0.2\\
   &  & CaII (H) & 4450 & - & -\\
   &  & G band & 4825 & 3.3 & 0.4\\
   &  & H$\gamma$ & 4873 & - & -\\
   &  & [OIII] & 5561 & +4.1 & 0.3\\
   &  & [OIII] & 5615 & +9.9 & 0.1\\
   &  & Mg & 5803 & 2.4 & 0.3\\
   &  & Na & 6607 & 2.0 & 0.1\\
   &  & [OI] & 7065 & +1.8 & 0.1\\
   &  & [NII] & 7343 & +2.4 & 0.3\\
   &  & H$\alpha$ & 7367 & +41 & 3\\
   &  & [NII] & 7383 & +4.3 & 0.6\\
   &  & [SII] & 7532 & +1.4 & -\\
   &  & [SII] & 7549 & +1.7 & -\\
   \hline
  3FGL J0059.1-5701 & 0.677 & MgII & 4694 & +46 & 4\\
   &  & FeII & $\sim$4976 & - & -\\
   &  & [NeV] & 5749 & - & -\\
   &  & [OII]d & 6257 & - & -\\
   &  & NeIII & 6494 & - & -\\
   &  & HeI & 6527 & +1.4 & 0.1\\
   &  & H$\epsilon$ & 6665 & +4.0 & 0.2\\
   &  & [SII] & 6873 & - & -\\
   &  & H$\delta$ & 6919 & - & -\\
   &  & H$\gamma$ & 7299 & +35 & 1\\

\hline\end{tabular}
\caption{The measured spectral features for our sample. In column 1 we report the 3FGL name, in column 2 its redshift as obtained from our optical spectra, in column 3 the line identification, in column 4 its observed wavelength in angstroms, and in columns 5 and 6 its observed equivalent width and its error. All unidentified lines are marked with a question mark, as well as all the unknown redshift values. Any given equivalent width values that could not be properly measured due to uncertainties are reported as null. When the measured equivalent width corresponds to an emission line, it is reported with a explicit $+$ sign; otherwise it corresponds to an absorption line. *:The $H\alpha$ and $H\beta$ lines present in the spectrum of 3FGL J1816.9-4944 are not attributed to the blazar, but to an intervening system.}
\end{center}
\end{table*}
   
\setcounter{table}{1}
   
\begin{table*}
\begin{center}
\begin{tabular}{|l|c|c|c|r|l|}
\hline\hline
  \multicolumn{1}{|c|}{Name} &
  \multicolumn{1}{c|}{z} &
  \multicolumn{1}{c|}{Line} &
  \multicolumn{1}{c|}{Obs. wavelength} &
  \multicolumn{1}{c|}{Eqw.} &
  \multicolumn{1}{c|}{Eqw. error} \\
  
  \multicolumn{1}{|c|}{[3FGL]} &
  \multicolumn{1}{c|}{} &
  \multicolumn{1}{c|}{} &
  \multicolumn{1}{c|}{[\AA ]} &
  \multicolumn{1}{c|}{[\AA ]} &
  \multicolumn{1}{c|}{[\AA ]} \\
\hline\hline
  3FGL J0115.8+2519 & ? & - & - & - & -\\
  \hline
  3FGL J0133.2-5259 & 0.925 & FeII & 5082 & - & -\\
   &  & FeII & 5095 & - & -\\
   &  & MgII & 5386 & +57 & 3\\
   &  & FeII & 5683 & - & -\\
   &  & [NeV] & 7173 & - & -\\
   &  & [NeIII] & 7445 & +4.1 & 0.4\\
   &  & HeI & 7485 & +5.1 & 0.4\\
   &  & H$\delta$ & 7892 & +18.1 & 0.9\\
   \hline
  3FGL J0211.2-0649 & 0.194 & CaII (K) & 4696 & 4.3 & 0.1\\
   &  & CaII (H) & 4737 & 3.0 & 0.1\\
   &  & G band & 5140 & 2.8 & 0.1\\
   &  & H$\beta$ & 5806 & 1.6 & 0.3\\
   &  & Mg & 6180 & 4.0 & 0.1\\
   \hline
  3FGL J0301.4-1652 & 0.278 & CaII (K) & 5023 & 4.0 & 0.3\\
   &  & CaII (H) & 5071 & 2.2 & 0.1\\
   &  & G band & 5500 & 2.8 & 0.4\\
   &  & [OIII] & 6335 & +2.3 & 0.2\\
   &  & [OIII] & 6398 & +1.9 & 0.2\\
   &  & MgI    & 6611 & 3.3 & 0.3 \\
   \hline
  3FGL J0540.4+5823 & ? & - & - & - & -\\
  \hline
  3FGL J0607.4+4739 & ? & - & - & - & -\\
  \hline
  3FGL J0623.3+3043 & ? & - & - & - & -\\
  \hline
  3FGL J0625.2+4440 & ? & - & - & - & -\\
  \hline
  3FGL J0653.6+2817 & ? & - & - & - & -\\
  \hline
  3FGL J0708.9+2239 & ? & - & - & - & -\\
  \hline
  3FGL J0723.7+2050 & ? & - & - & - & -\\
  \hline
  3FGL J0748.8+4929 & ? & - & - & - & -\\
  \hline
  3FGL J0922.8-3959 & 0.595 & MgII & 4456 & +65 & 4\\
   &  & H$\gamma$ & 6534 & +15 & 1\\
   &  & H$\delta$ & 6930 & +39 & 3\\
   &  & H$\beta$ & 7745 & +97 & 5\\
   &  & [OIII] & +6.9 & 1 \\
   &  & [OIII] & +14 & 2 \\
   \hline
  3FGL J0952.8+0711 & ? & - & - & - & -\\
   \hline
  3FGL J1052.8-3741 & ? & - & - & - & -\\
    \hline
  3FGL J1440.0-3955 & 0.300 & CaII (K) & 5115 & 5.8 & 0.2\\
   &  & CaII (H) & 5160 & 4.4 & 0.1\\
   &  & G band & 5595 & 3.0 & 0.2\\
   &  & MgI & 6728 & 2.8 & 0.1 \\
   \hline
  3FGL J1559.8-2525 & 0.416 & CaII (K) & 5579 & 4.6 & 0.3\\
   &  & CaII (H) & 5620 & 2.3 & 0.1\\
\hline\end{tabular}
\caption{(Continued)}
\end{center}
\end{table*}

\setcounter{table}{1}

\begin{table*}
\begin{center}
\begin{tabular}{|l|c|c|c|r|l|}
\hline\hline
  \multicolumn{1}{|c|}{Name} &
  \multicolumn{1}{c|}{z} &
  \multicolumn{1}{c|}{Line} &
  \multicolumn{1}{c|}{Obs. wavelength} &
  \multicolumn{1}{c|}{Eqw.} &
  \multicolumn{1}{c|}{Eqw. error} \\
  
  \multicolumn{1}{|c|}{[3FGL]} &
  \multicolumn{1}{c|}{} &
  \multicolumn{1}{c|}{} &
  \multicolumn{1}{c|}{[\AA ]} &
  \multicolumn{1}{c|}{[\AA ]} &
  \multicolumn{1}{c|}{[\AA ]} \\
\hline\hline
   \hline
  3FGL J1816.9-4944 & 1.70 & CIV & 4183 & +49 & 4\\
   &  & H$\beta$* & 4860 & 2.1 & 0.1\\
   &  & H$\alpha$* & 6560 & 2.7 & 0.1\\
   &  & MgII & 7562 & +3.9 & 0.2\\
   \hline
  3FGL J1842.3-5841 & 0.421 & CaII (K) & 5592 & 2.6 & 0.4\\
   &  & CaII (H) & 5641 & 2.2 & 0.2\\
   &  & G band & 6117 & 3.4 & 0.5\\
      \hline
  3FGL J1911.4-1908 & 0.138 & CaII (K) & 4476 & 4.0 & 0.3\\
   &  & CaII (H) & 4514 & 3.2 & 0.2\\
   &  & G band & 4894 & 3.3 & 0.3\\
   &  & H$\beta$ & 5542 & - & -\\
   &  & [OIII] & 5636 & +1.8 & 0.2\\
   &  & [OIII] & 5695 & +0.9 & 0.1\\
   &  & Na & 6705 & 3.0 & 0.1\\
   &  & [NII] & 7451 & 1.0 & 0.1\\
   &  & H$\alpha$ & 7567 & +0.7 & 0.3\\
   &  & [NII] & 7491 & 2.8 & 0.1\\
   \hline
  3FGL J1954.9-5640 & 0.221 & CaII (K) & 4802 & 1.6 & 0.2\\
   &  & CaII (H) & 4846 & 1.5 & 0.1\\
   &  & G band & 5255 & 0.9 & 0.1\\
   &  & MgI+Tell. & 6316 & 2.5 & 0.2\\
   \hline
  3FGL J2046.7-1011 & - & - & - & - & -\\
   \hline
  3FGL J2103.9-6233 & ? & - & - & - & -\\
  \hline
  3FGL J2107.7-4822 & $\ge$0.519 & MgII & 4246 & 2.3 & 0.2\\
   &  & MgII & 4258 & 2.0 & 0.1\\
   &  & ? & 7085 & +26.2 & 2.0\\
   \hline
  3FGL J2126.5-3926 & ? & - & - & - & -\\
  \hline
  3FGL J2144.2+3132 & ? & - & - & - & -\\
  \hline
  3FGL J2159.2-2841 & 0.270 & CaII (K) & 4999 & 2.6 & 0.1\\
   &  & CaII (H) & 5042 & 2.9 & 0.1\\
   &  & G band & 5463 & 2.2 & 0.2\\
   \hline
  3FGL J2212.6+2801 & ? & - &  & 0.0 & 0.0\\
  \hline
  3FGL J2213.6-4755 & 1.529 & FeII & 4089 & 3.5 & 0.1\\
   &  & MgII & 7071 & 2.2 & 0.4\\
   &  & MgII & 7089 & 1.7 & 0.2\\
   \hline
  3FGL J2305.3-4219 & ? & - &  & - & -\\
  \hline
  3FGL J2316.8-5209 & 0.646 & MnII? & 4241 & - & -\\
   &  & FeII? & 4253 & - & -\\
   &  & CaII (K) & 6477 & 1.2 & 0.1\\
   &  & CaII (H) & 6529 & 1.4 & 0.1\\
   \hline
  3FGL J2348.4-5100 & 0.398 & CaII (K) & 5503 & 1.5 & 0.1\\
   &  & CaII (H) & 5549 & 0.8 & 0.1\\
   &  & G band & 6013 & 1.6 & 0.2\\
\hline\end{tabular}
\caption{(Continued)}
\end{center}
\end{table*}

\section{Summary and conclusions}

This is paper number VIII of the series dedicated to the optical spectroscopic campaign of UGSs and BCUs listed in the \fer\ catalogs, aiming to obtain spectroscopic confirmation of their associated and/or potential low-energy counterparts. The analysis presented here is almost entirely dedicated to BCUs as occurred in additional works \citep[see e.g.,][]{optbcu}. 

Here we reported the follow up spectroscopic observations of a sample listing 36 targets, carried out thanks to the SPM, SOAR, KPNO and NOT observatories. All of them were observed between January 2016 and August 2017. The selected sample includes 30 BCUs, 4 BZBs already classified in the latest version of the Roma-BZCAT and associated with \fer\ sources but lacking a $z$ estimate, and 1 BL Lac candidate also associated with a \fer\ source but lacking a spectroscopic confirmation of its classification. Additionally, we re-observed the object VSS\,J095249+071330, an infrared source lying within the position uncertainty region of the UGS 3FGL\,J0952.8+0711 showing IR colors similar to those of known \fer\ blazars \citep{strip}, and already identified as a BL Lac object \citep{Paiano17d}.

Our results can be summarized as follows.

\begin{itemize}
\item Twenty-three out of thirty BCUs are classifiable as BL Lac objects. The remaining 7 are all QSOs, and thus presenting a flat radio spectrum can be classified as BZQs, with the only exception of PMN\,J0049-4457, associated with 3FGL\,J0050.0-4458, for which the optical spectrum resembles that of an absorbed Seyfert galaxy.
\item For 12 of the BL Lac objects we are able to obtain a $z$ estimate, one of them being only a lower limit due to spectral features ascribable to intervening systems along the line of sight.
\item All the four BZBs re-observed to get an estimate of their redshift show featureless optical spectra. This is also the case of 5BZB\,J0607+4739, the BL Lac candidate listed in the Roma-BZCAT and associated with 3FGL\,J0607.4+4739. Our observations confirm its BL Lac nature. 
\item Finally, we also re-observed the BZB object VSS\,J095249+071330, already associated with 3FGL J0952.8+0711. We do not detect any spectral variability with the results published previously in the literature.
\item We could associate an optical counterpart for the BCU 3FGL J2213.6-4755, which is SUMSS J221330-475426. This source is a BL Lac lying at a redshift of $>$1.529, which means it is one of the rare, most distant BL Lacs, which are counted by tens.
\end{itemize}

The current analysis increases the number of optical spectroscopic confirmation for UGSs and BCUs by $\sim$20\% \citep{quest}. It also confirms the trends that a large fraction of the sources classified tend to be BL Lacs, regardless of whether they were selected from the BCU list of the \fer\ catalogs, or among the objects lying within the positional uncertainty regions of UGSs. These objects are not only the largest known population of gamma-ray sources but continue to be the most elusive class in the gamma-ray sky.

This paper reports the results achieved with single-slit spectroscopic observations collected until August 2017. More results are to come in the near future, given that 10 more nights will be used to complete the BCU observations in the Southern Hemisphere. We expect to carry on with our campaign until the end of 2019, since more nights have been awarded in the next semesters. 

\newpage
\begin{onecolumn}
\centering
 \begin{figure}[t]
\includegraphics[scale=1.0]{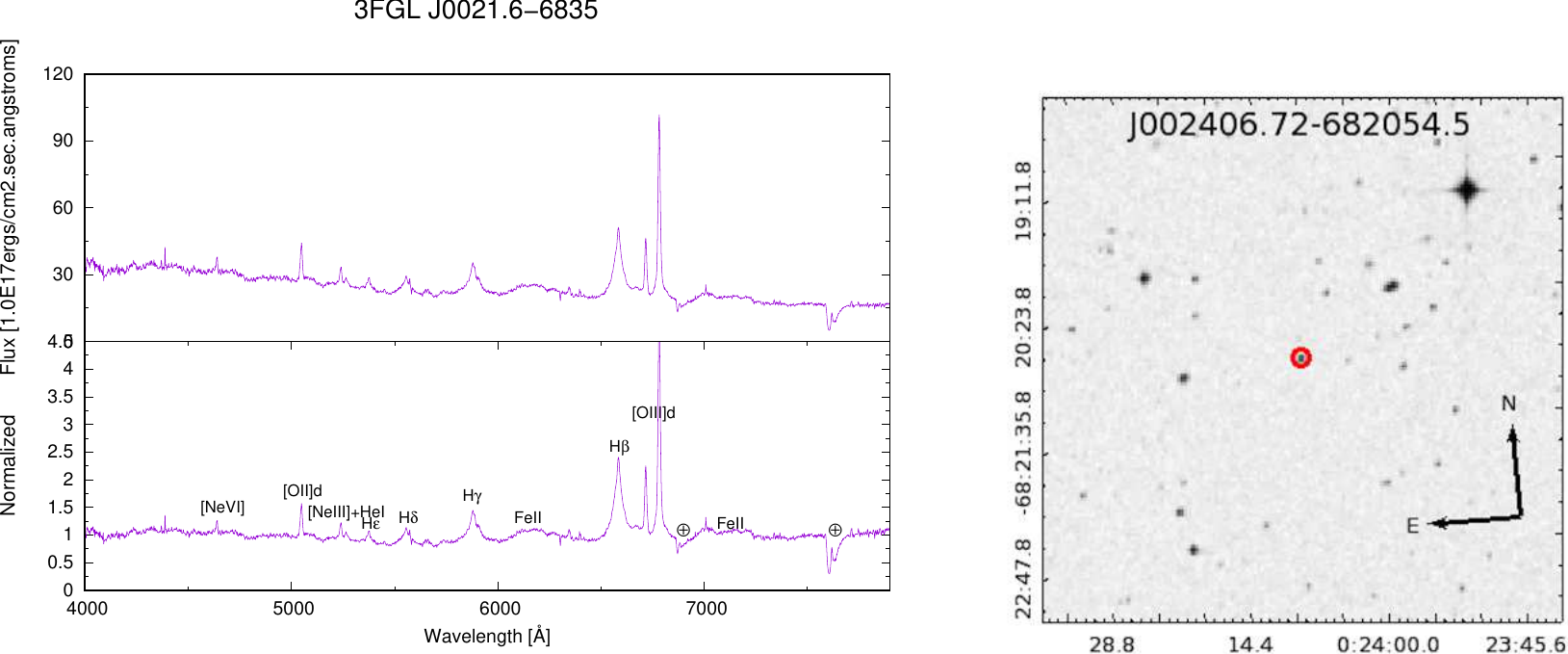}
 \caption{(Left panel) Top: Optical spectrum of WISE\,J002406.72-682054.5 associated with 3FGL\,J0021.6-6835. Bottom: The same spectrum, normalised to highlight features (if any). If present, telluric lines are marked with $\oplus$, and features due to contamination from diffuse interstellar bands are marked with DIB. If there are any doublets, these are marked with a d. The absorption line at $\sim$5890\AA\, which is NaI from the Milky Way, is marked as NaIMW. Unidentified lines are marked with a question mark. (Right panel) The finding chart ( 5'$\times$ 5' ) retrieved from the Digitized Sky Survey highlighting the location of the optical source: WISE\,J002406.72-682054.5 (red circle)} 
 \label{fig:1}
 \end{figure}
 \centering
 \begin{figure}[b]
 \includegraphics[scale=1.0]{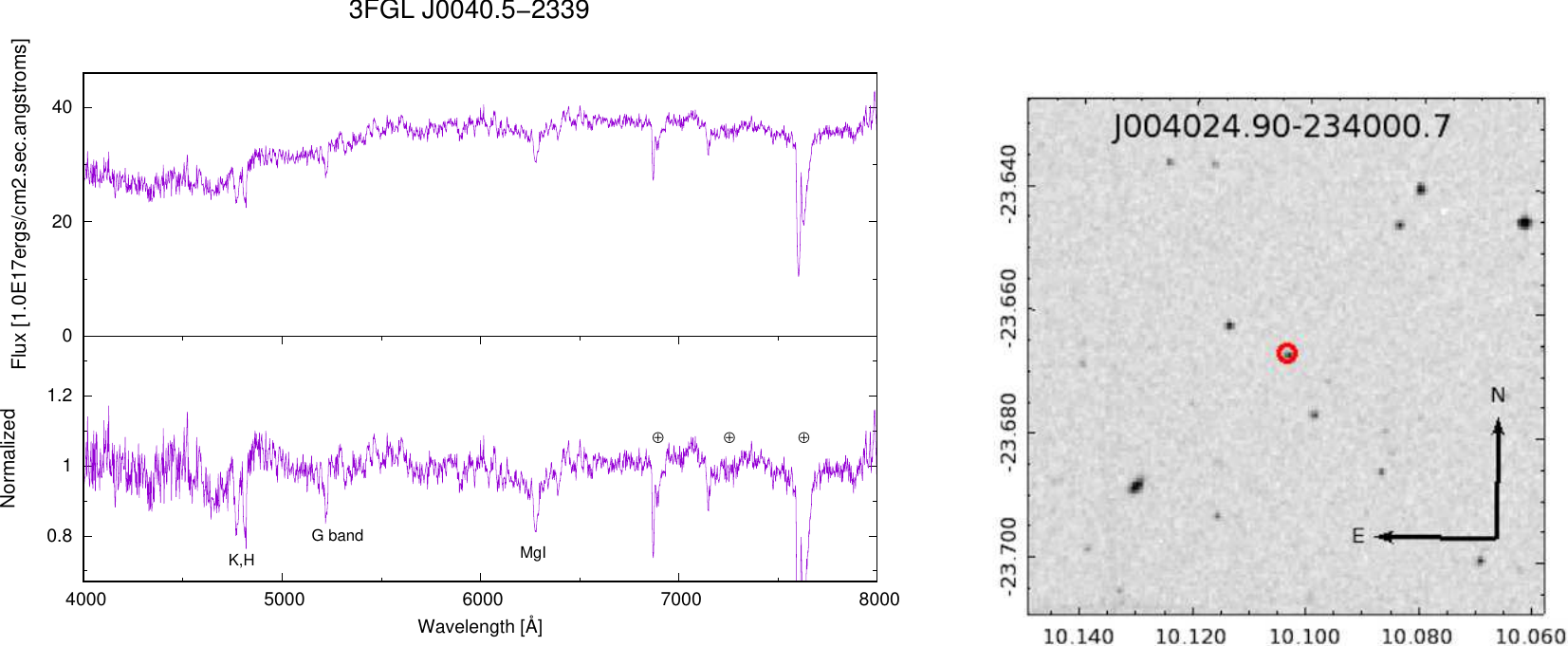}
 \caption{(Left panel) Top: Optical spectrum of WISE\,J004024.90-234000.7 associated with 3FGL\,J0040.5-2339. Bottom: The same spectrum, normalised to highlight features (if any). If present, telluric lines are marked with $\oplus$, and features due to contamination from diffuse interstellar bands are marked with DIB. If there are any doublets, these are marked with a d. The absorption line at $\sim$5890\AA\, which is NaI from the Milky Way, is marked as NaIMW. Unidentified lines are marked with a question mark. (Right panel)  The finding chart ( 5'$\times$ 5' ) retrieved from the Digitized Sky Survey highlighting the location of the optical source: WISE\,J002406.72-682054.5 (red circle)} 
 \label{fig:2}
 \end{figure}
 
  \end{onecolumn}
 \newpage
\begin{onecolumn}

\centering
 \begin{figure}[t]
 \includegraphics[scale=1.0]{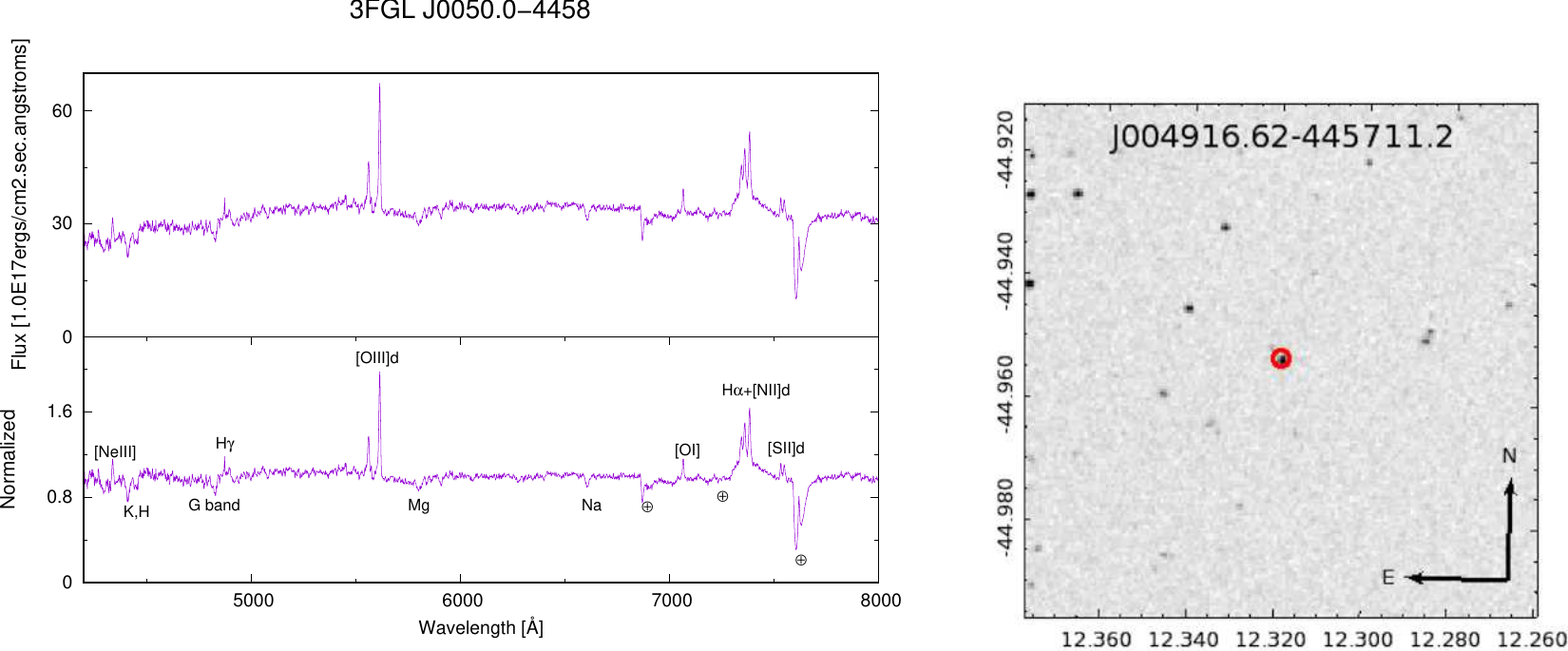}
 \caption{(Left panel) Top: Optical spectrum of WISE\,J004916.62-445711.2 associated with 3FGL\,J0050.0-4458. Bottom: The same spectrum, normalised to highlight features (if any). If present, telluric lines are marked with $\oplus$, and features due to contamination from diffuse interstellar bands are marked with DIB. If there are any doublets, these are marked with a d. The absorption line at $\sim$5890\AA\, which is NaI from the Milky Way, is marked as NaIMW. Unidentified lines are marked with a question mark. (Right panel)  The finding chart ( 5'$\times$ 5' ) retrieved from the Digitized Sky Survey highlighting the location of the optical source: WISE\,J004916.62-445711.2 (red circle)} 
 \label{fig:3}
 \end{figure}
\centering
 \begin{figure}[b]
 \includegraphics[scale=1.0]{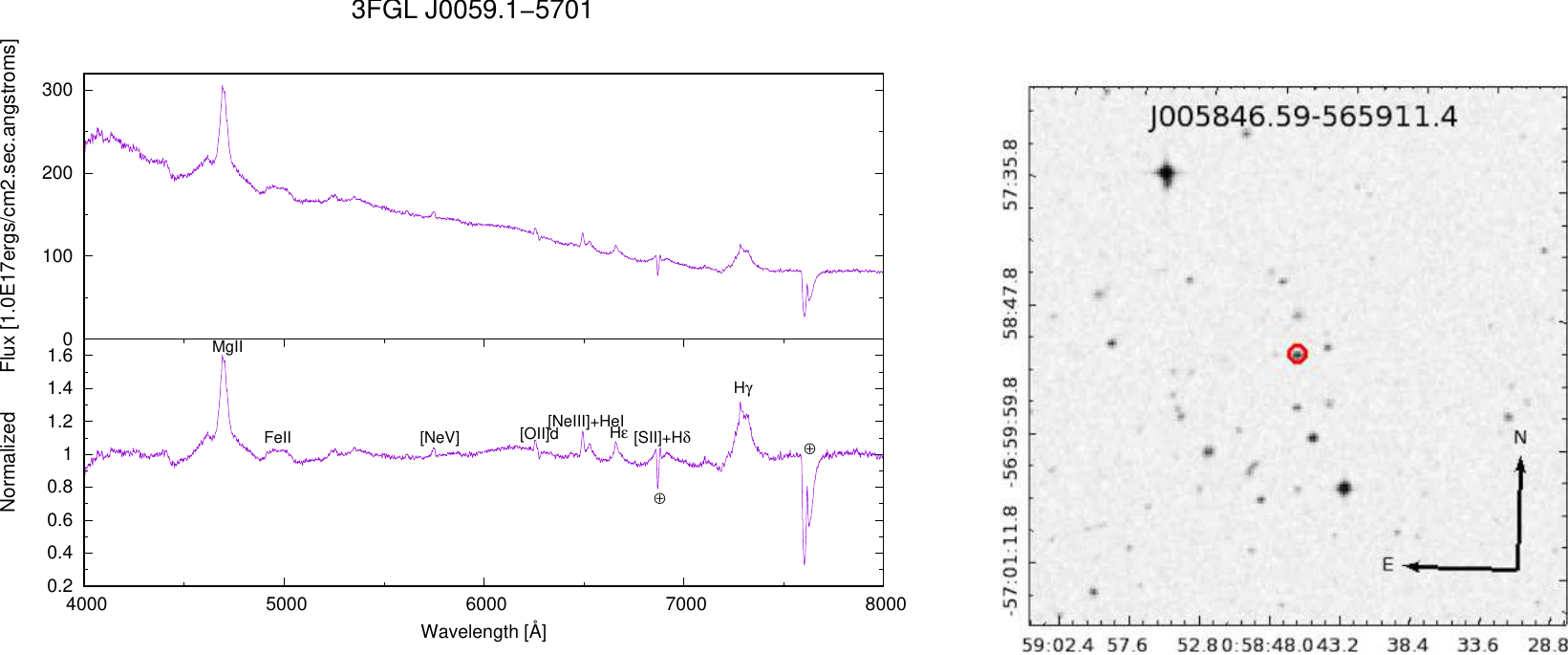}
 \caption{(Left panel) Top: Optical spectrum of WISE\,J005846.59-565911.4 associated with 3FGL\,J0059.1-5701. Bottom: The same spectrum, normalised to highlight features (if any). If present, telluric lines are marked with $\oplus$, and features due to contamination from diffuse interstellar bands are marked with DIB. If there are any doublets, these are marked with a d. The absorption line at $\sim$5890\AA\, which is NaI from the Milky Way, is marked as NaIMW. Unidentified lines are marked with a question mark. (Right panel)  The finding chart ( 5'$\times$ 5' ) retrieved from the Digitized Sky Survey highlighting the location of the optical source: WISE\,J005846.59-565911.4 (red circle)} 
 \label{fig:4}
 \end{figure}
 
   \end{onecolumn}
 \newpage
 \begin{onecolumn}

 \centering
 \begin{figure}[t]
 \includegraphics[scale=1.0]{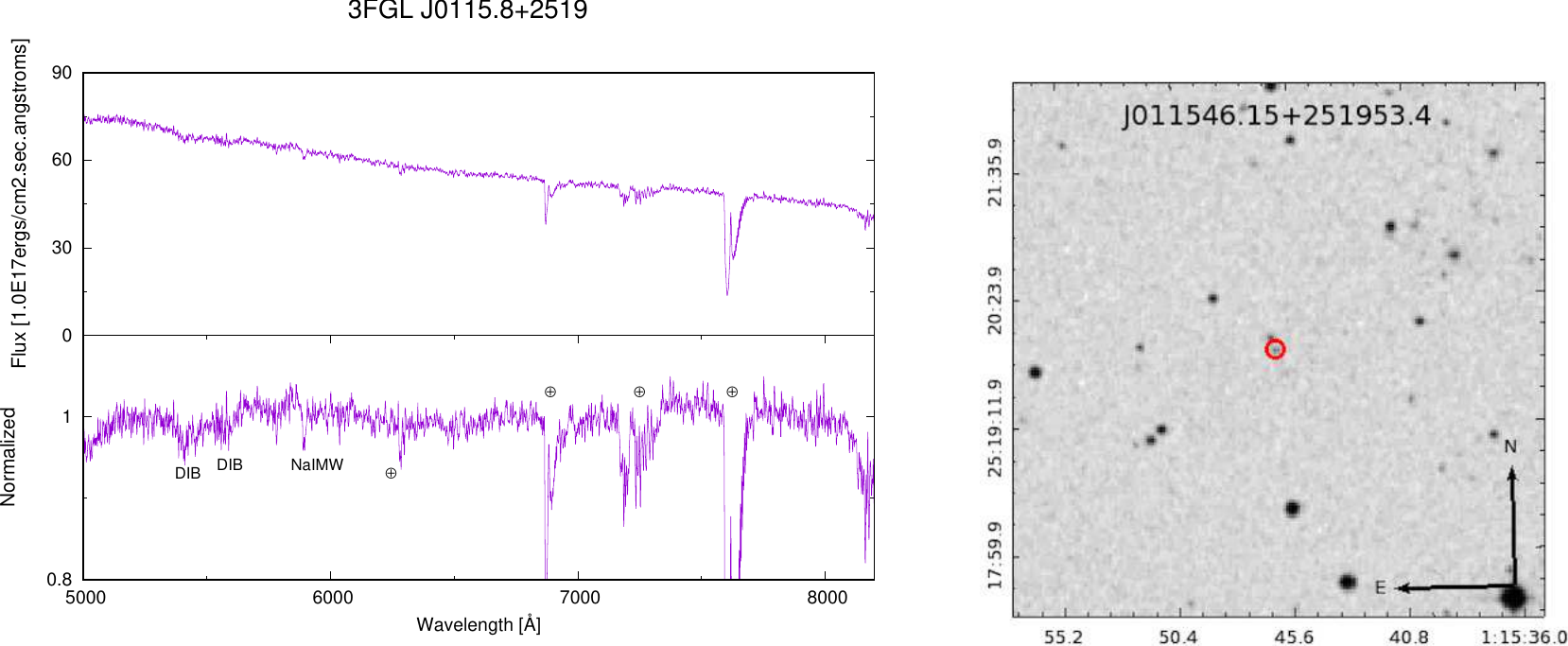}
 \caption{(Left panel) Top: Optical spectrum of WISE\,J011546.15+251953.4 associated with 3FGL\,J0115.8+2519. Bottom: The same spectrum, normalised to highlight features (if any). If present, telluric lines are marked with $\oplus$, and features due to contamination from diffuse interstellar bands are marked with DIB. If there are any doublets, these are marked with a d. The absorption line at $\sim$5890\AA\, which is NaI from the Milky Way, is marked as NaIMW. Unidentified lines are marked with a question mark. (Right panel)  The finding chart ( 5'$\times$ 5' ) retrieved from the Digitized Sky Survey highlighting the location of the optical source: WISE\,J011546.15+251953.4 (red circle)} 
 \label{fig:5}
 \end{figure}
 \centering
 \begin{figure}[b]
 \includegraphics[scale=1.0]{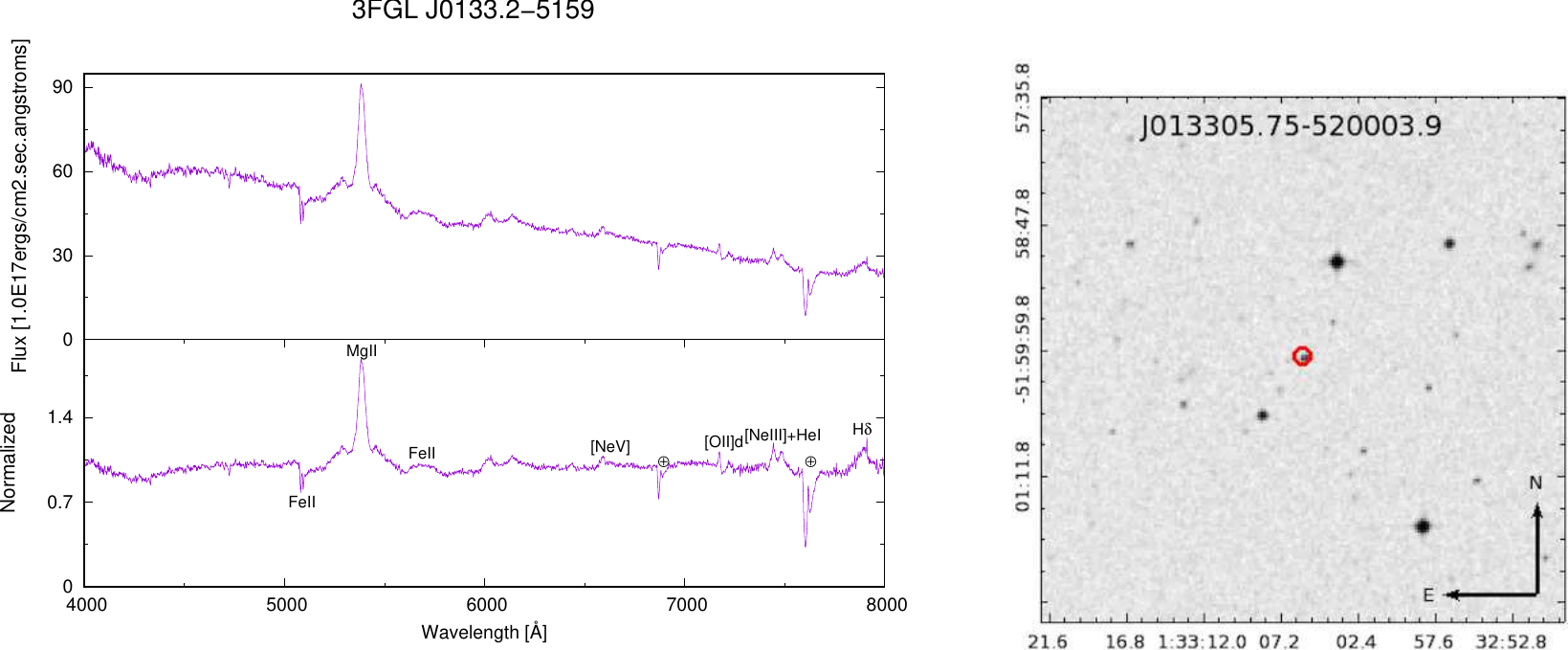}
 \caption{(Left panel) Top: Optical spectrum of WISE\,J013305.75-520003.9 associated with 3FGL\,J0133.2-5159. Bottom: The same spectrum, normalised to highlight features (if any). If present, telluric lines are marked with $\oplus$, and features due to contamination from diffuse interstellar bands are marked with DIB. If there are any doublets, these are marked with a d. The absorption line at $\sim$5890\AA\, which is NaI from the Milky Way, is marked as NaIMW. Unidentified lines are marked with a question mark. (Right panel)  The finding chart ( 5'$\times$ 5' ) retrieved from the Digitized Sky Survey highlighting the location of the optical source: WISE\,J013305.75-520003.9 (red circle)} 
 \label{fig:6}
 \end{figure}
 
    \end{onecolumn}
 \newpage
 \begin{onecolumn}
 
 \centering
 \begin{figure}[t]
 \includegraphics[scale=1.0]{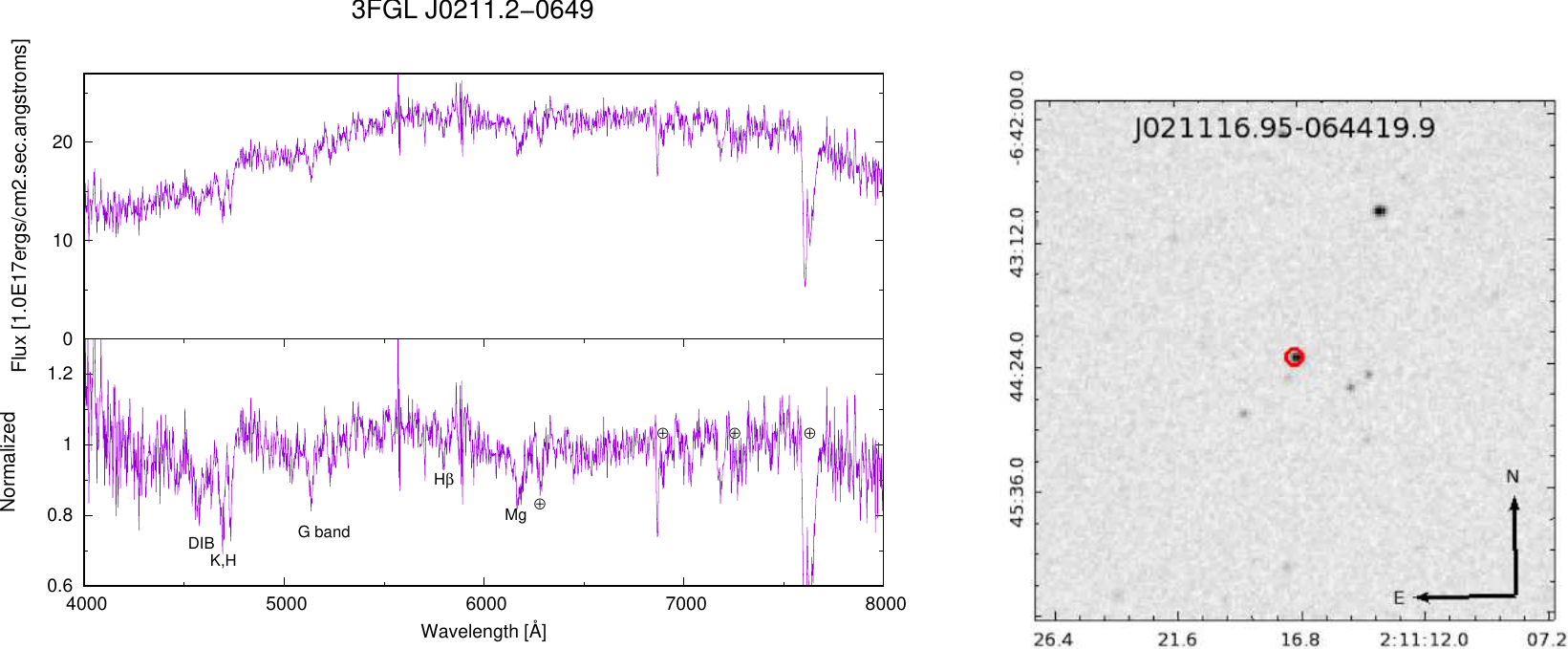}
 \caption{(Left panel) Top: Optical spectrum of WISE\,J021116.95-064419.9 associated with 3FGL\,J0211.0+1922. Bottom: The same spectrum, normalised to highlight features (if any). If present, telluric lines are marked with $\oplus$, and features due to contamination from diffuse interstellar bands are marked with DIB. If there are any doublets, these are marked with a d. The absorption line at $\sim$5890\AA\, which is NaI from the Milky Way, is marked as NaIMW. Unidentified lines are marked with a question mark. (Right panel)  The finding chart ( 5'$\times$ 5' ) retrieved from the Digitized Sky Survey highlighting the location of the optical source: WISE\,J021116.95-064419.9 (red circle)} 
 \label{fig:7}
 \end{figure}
 \centering
 \begin{figure}[b]
 \includegraphics[scale=1.0]{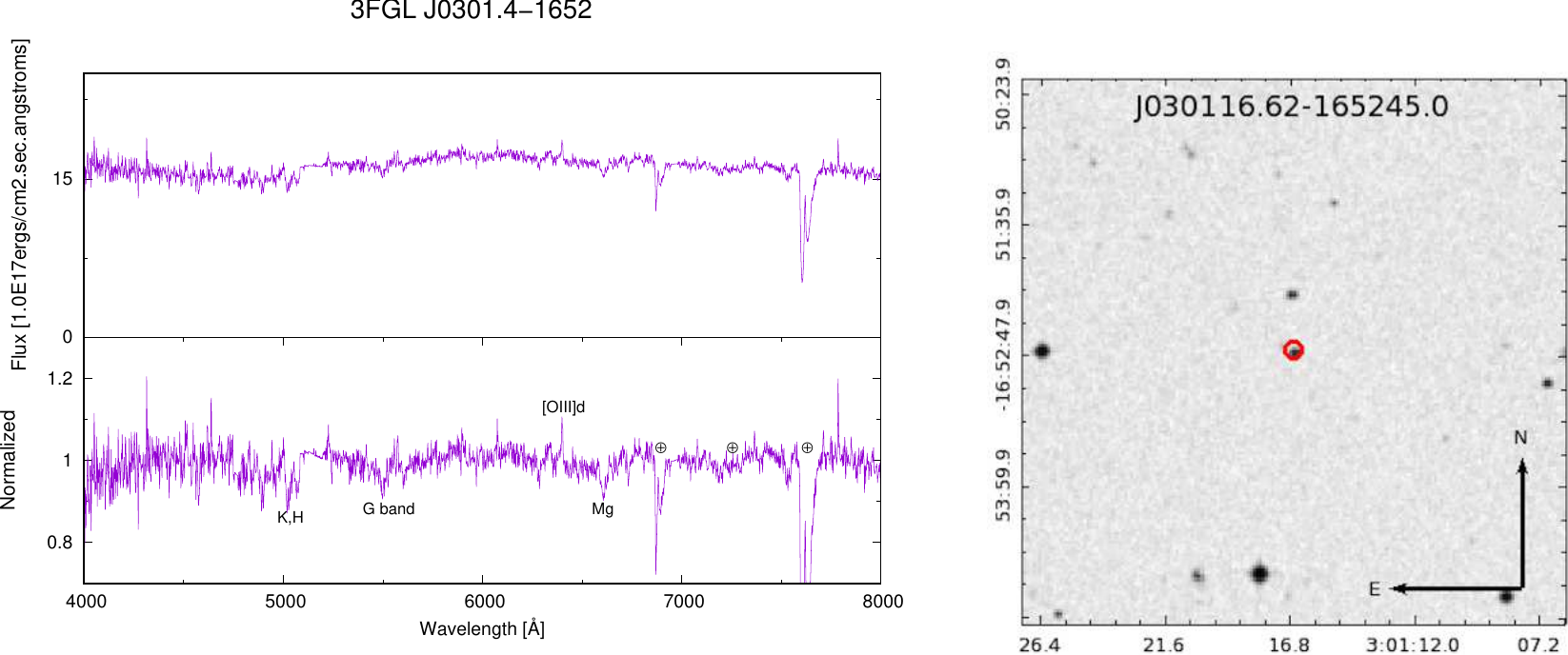}
 \caption{(Left panel) Top: Optical spectrum of WISE\,J030116.62-165245.0 associated with 3FGL\,J0301.4-1652. Bottom: The same spectrum, normalised to highlight features (if any). If present, telluric lines are marked with $\oplus$, and features due to contamination from diffuse interstellar bands are marked with DIB. If there are any doublets, these are marked with a d. The absorption line at $\sim$5890\AA\, which is NaI from the Milky Way, is marked as NaIMW. Unidentified lines are marked with a question mark. (Right panel)  The finding chart ( 5'$\times$ 5' ) retrieved from the Digitized Sky Survey highlighting the location of the optical source: WISE\,J030116.62-165245.0 (red circle)} 
 \label{fig:8}
 \end{figure}
  
    \end{onecolumn}
 \newpage
 \begin{onecolumn}
 
 \centering
 \begin{figure}[t]
 \includegraphics[scale=1.0]{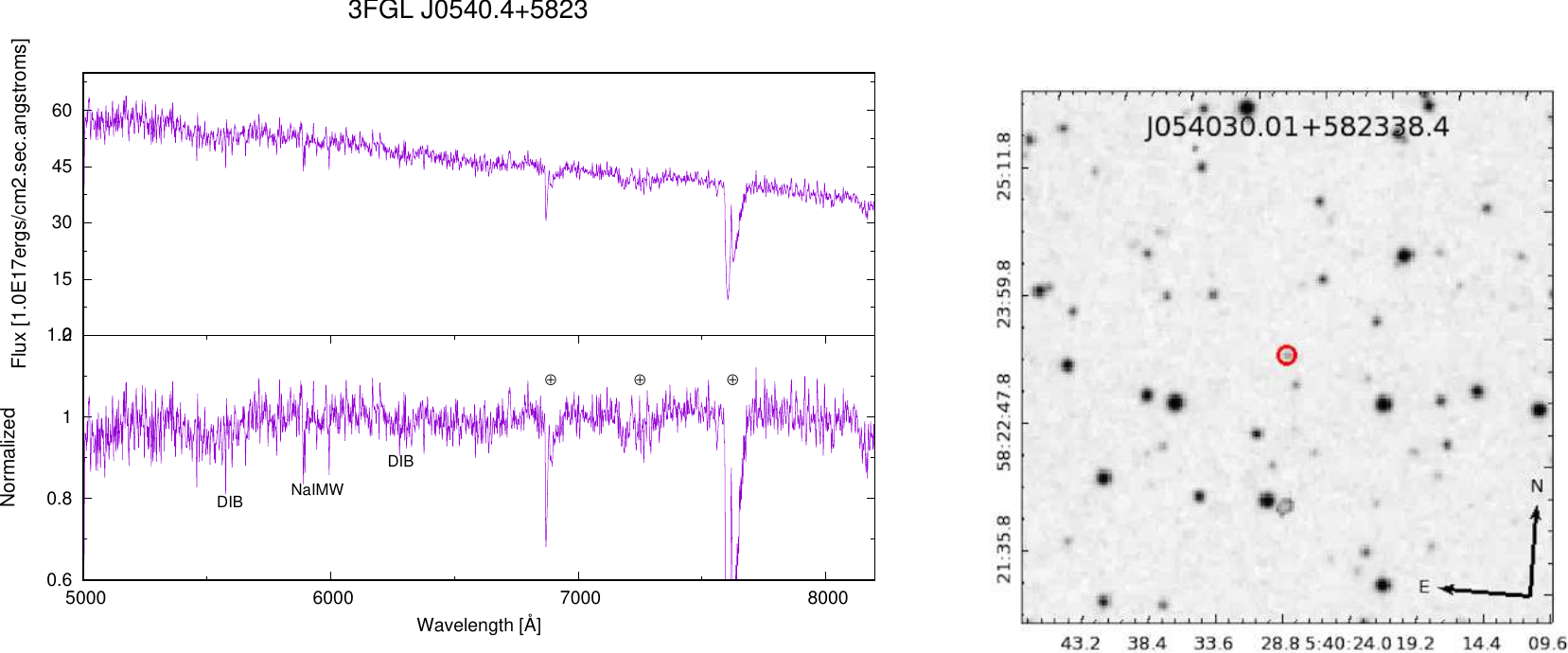}
 \caption{(Left panel) Top: Optical spectrum of WISE\,J054030.01+582338.4 associated with 3FGL\,J0540.4+5823. Bottom: The same spectrum, normalised to highlight features (if any). If present, telluric lines are marked with $\oplus$, and features due to contamination from diffuse interstellar bands are marked with DIB. If there are any doublets, these are marked with a d. The absorption line at $\sim$5890\AA\, which is NaI from the Milky Way, is marked as NaIMW. Unidentified lines are marked with a question mark. (Right panel)  The finding chart ( 5'$\times$ 5' ) retrieved from the Digitized Sky Survey highlighting the location of the optical source: WISE\,J054030.01+582338.4 (red circle)} 
 \label{fig:9}
 \end{figure}
 \centering
 \begin{figure}[b]
 \includegraphics[scale=1.0]{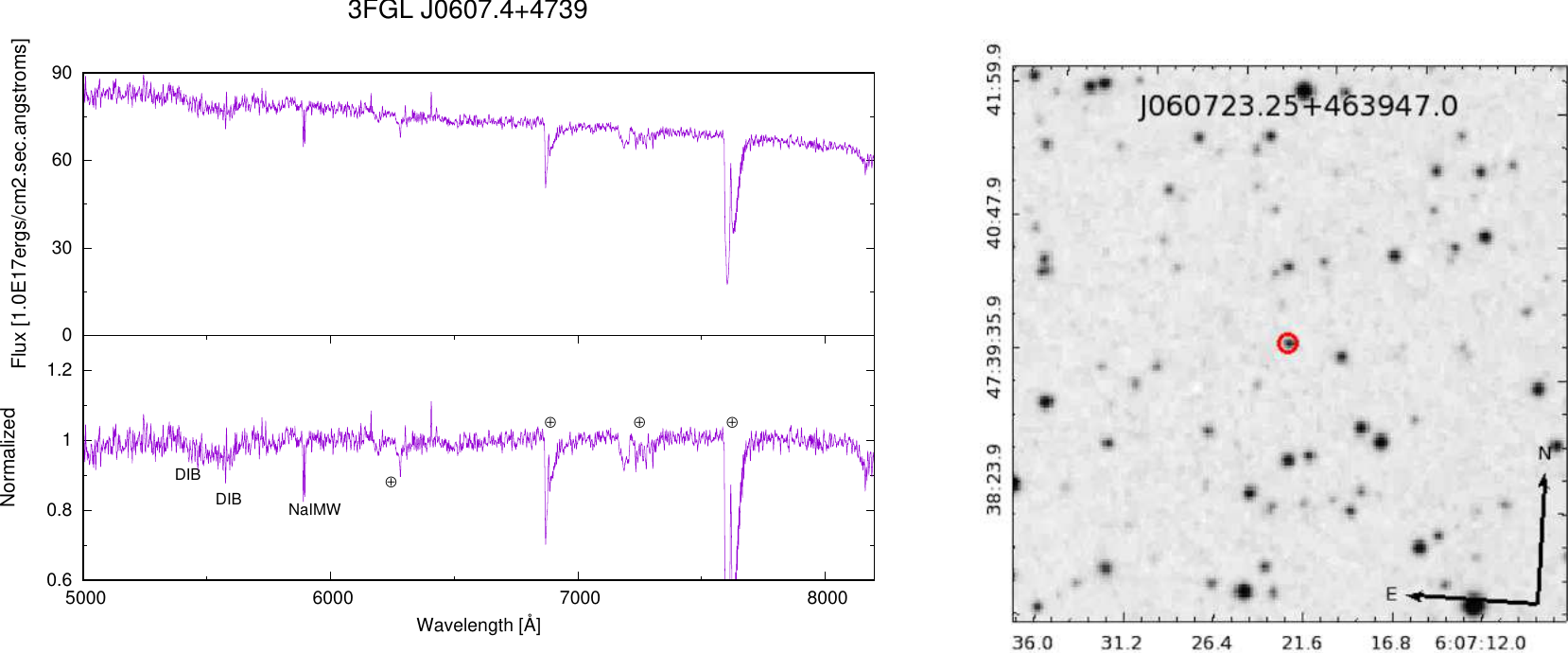}
 \caption{(Left panel) Top: Optical spectrum of WISE\,J060723.25+463947.0 associated with 3FGL\,J0607.4+4739. Bottom: The same spectrum, normalised to highlight features (if any). If present, telluric lines are marked with $\oplus$, and features due to contamination from diffuse interstellar bands are marked with DIB. If there are any doublets, these are marked with a d. The absorption line at $\sim$5890\AA\, which is NaI from the Milky Way, is marked as NaIMW. Unidentified lines are marked with a question mark. (Right panel)  The finding chart ( 5'$\times$ 5' ) retrieved from the Digitized Sky Survey highlighting the location of the optical source: WISE\,J060723.25+463947.0 (red circle)} 
 \label{fig:10}
 \end{figure}
  
    \end{onecolumn}
 \newpage
 \begin{onecolumn}
 
 \centering
 \begin{figure}[t]
 \includegraphics[scale=1.0]{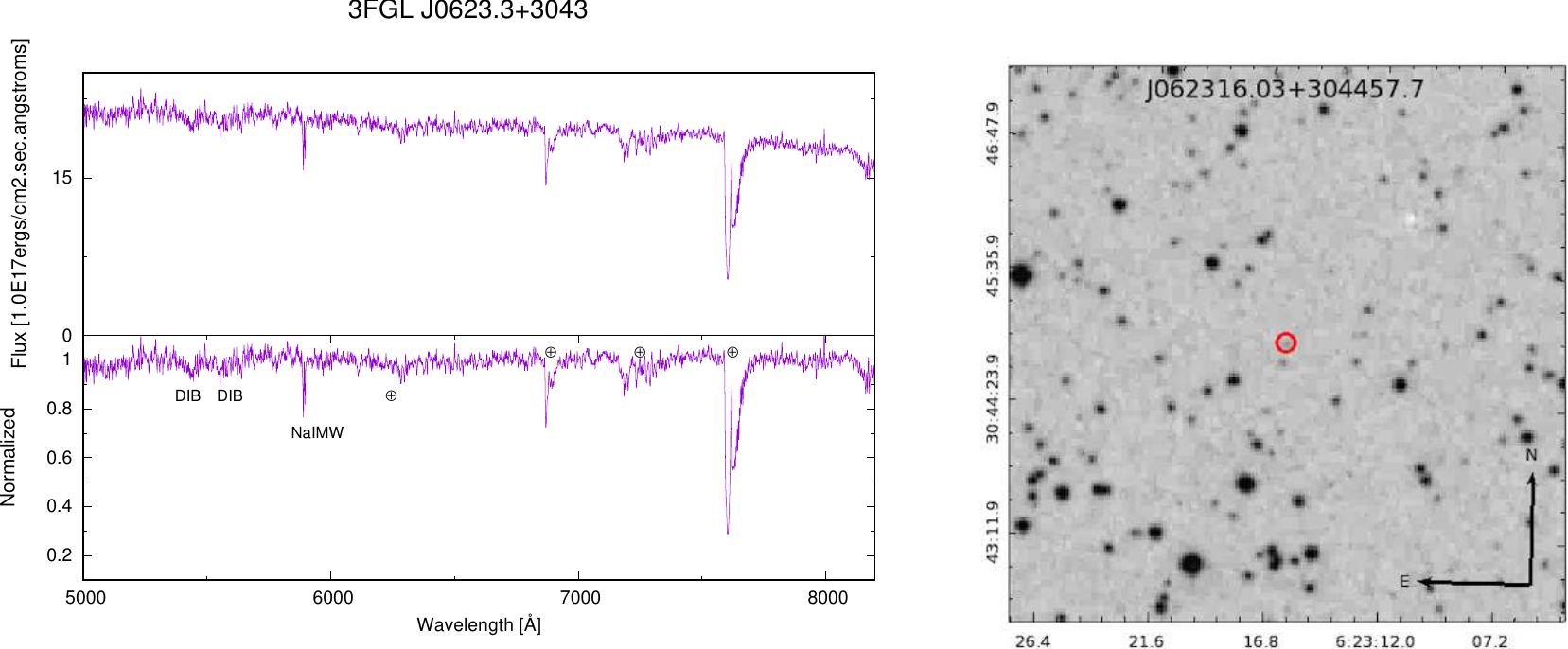}
 \caption{(Left panel) Top: Optical spectrum of WISE\,J062316.03+304457.7 associated with 3FGL\,J0623.3+3043. Bottom: The same spectrum, normalised to highlight features (if any). If present, telluric lines are marked with $\oplus$, and features due to contamination from diffuse interstellar bands are marked with DIB. If there are any doublets, these are marked with a d. The absorption line at $\sim$5890\AA\, which is NaI from the Milky Way, is marked as NaIMW. Unidentified lines are marked with a question mark. (Right panel)  The finding chart ( 5'$\times$ 5' ) retrieved from the Digitized Sky Survey highlighting the location of the optical source: WISE\,J062316.03+304457.7 (red circle)} 
 \label{fig:11}
 \end{figure}
 \centering
 \begin{figure}[b]
 \includegraphics[scale=1.0]{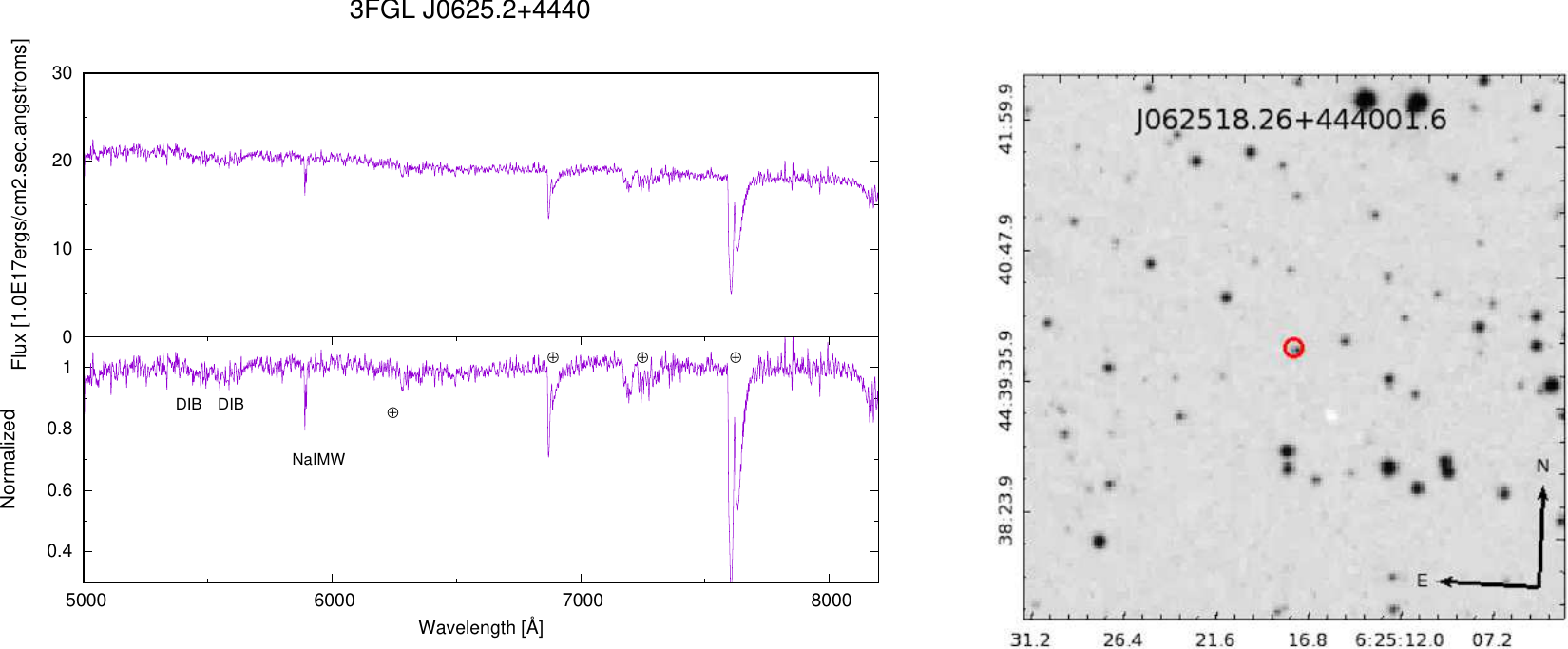}
 \caption{(Left panel) Top: Optical spectrum of WISE\,J062518.26+444001.6 associated with 3FGL\,J0625.2+4440. Bottom: The same spectrum, normalised to highlight features (if any). If present, telluric lines are marked with $\oplus$, and features due to contamination from diffuse interstellar bands are marked with DIB. If there are any doublets, these are marked with a d. The absorption line at $\sim$5890\AA\, which is NaI from the Milky Way, is marked as NaIMW. Unidentified lines are marked with a question mark. (Right panel)  The finding chart ( 5'$\times$ 5' ) retrieved from the Digitized Sky Survey highlighting the location of the optical source: WISE\,J062518.26+444001.6 (red circle)} 
 \label{fig:12}
 \end{figure}
  
    \end{onecolumn}
 \newpage
 \begin{onecolumn}
 
 \centering
 \begin{figure}[t]
 \includegraphics[scale=1.0]{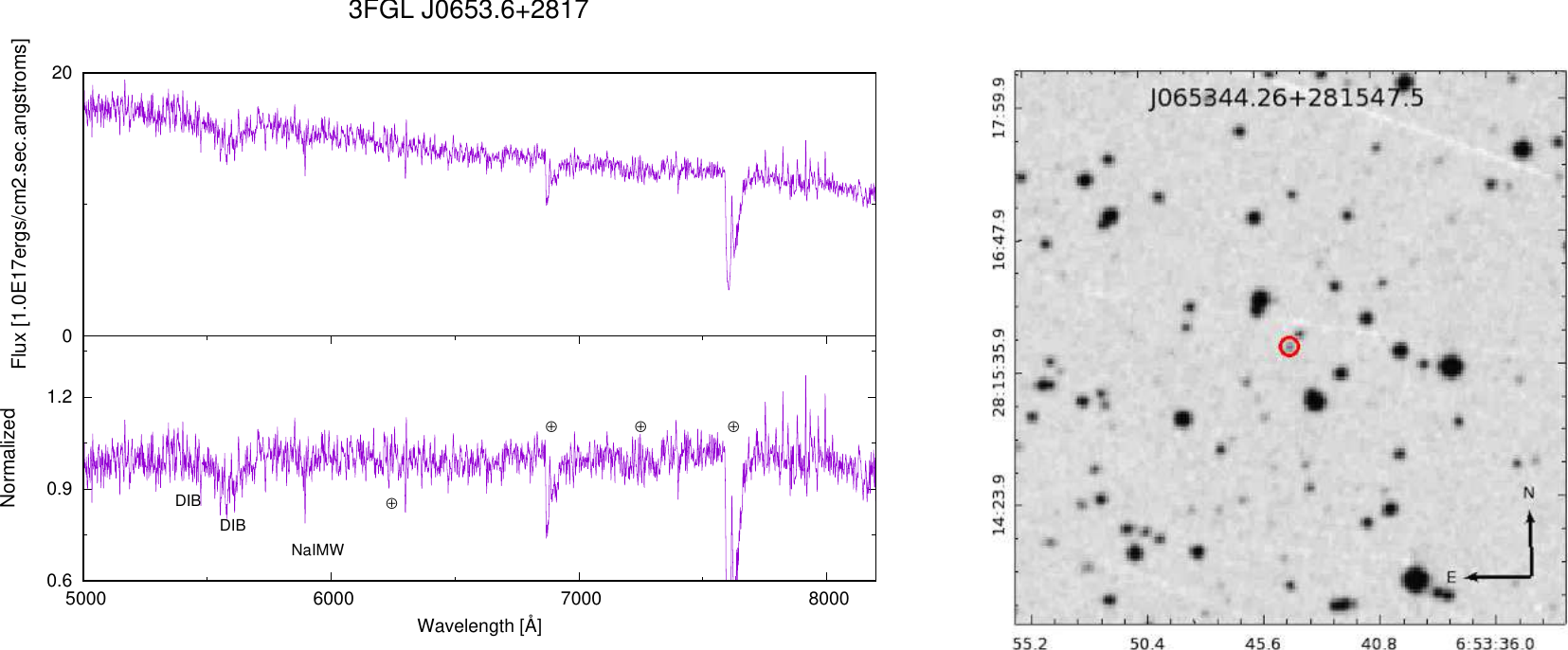}
 \caption{(Left panel) Top: Optical spectrum of WISE\,J065344.26+281547.5 associated with 3FGL\,J0653.6+2817. Bottom: The same spectrum, normalised to highlight features (if any). If present, telluric lines are marked with $\oplus$, and features due to contamination from diffuse interstellar bands are marked with DIB. If there are any doublets, these are marked with a d. The absorption line at $\sim$5890\AA\, which is NaI from the Milky Way, is marked as NaIMW. Unidentified lines are marked with a question mark. (Right panel)  The finding chart ( 5'$\times$ 5' ) retrieved from the Digitized Sky Survey highlighting the location of the optical source: WISE\,J065344.26+281547.5 (red circle)} 
 \label{fig:13}
 \end{figure}
 \centering
 \begin{figure}[b]
 \includegraphics[scale=1.0]{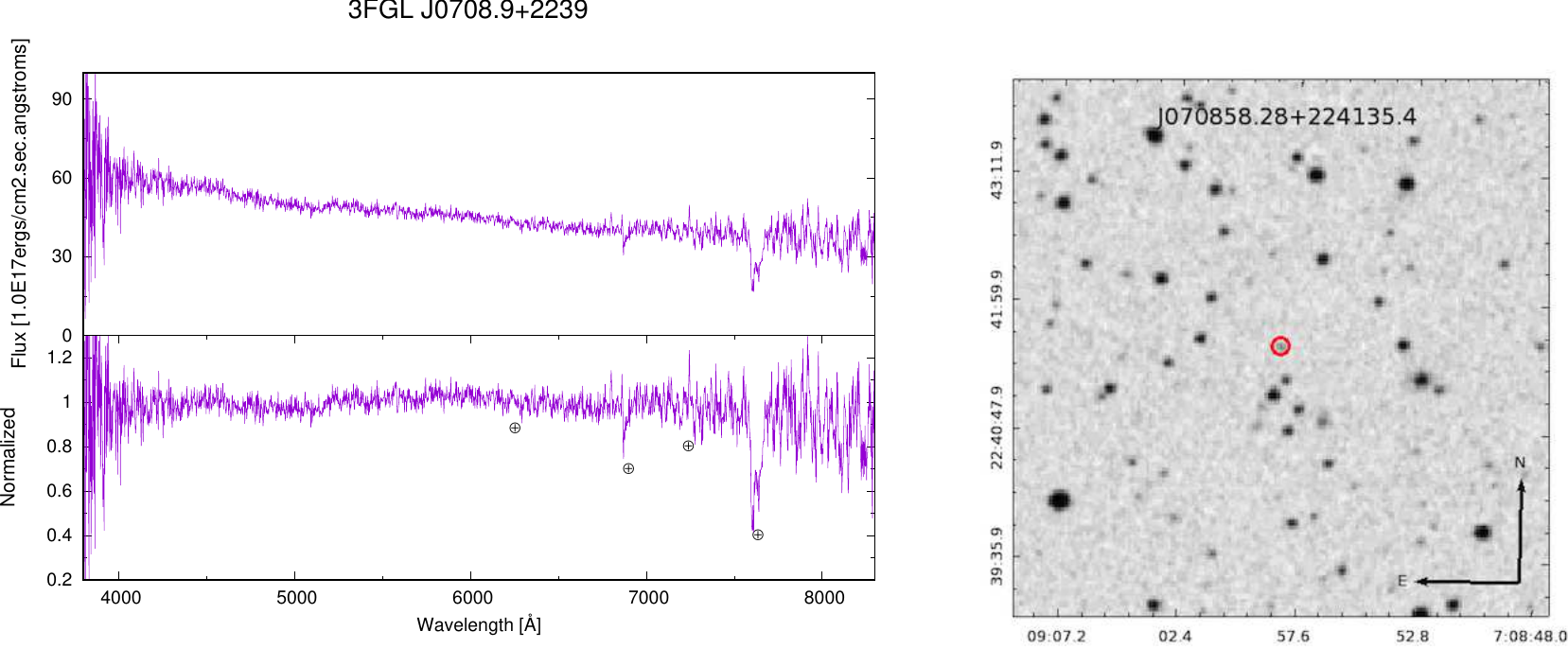}
 \caption{(Left panel) Top: Optical spectrum of WISE\,J070858.28+224135.4 associated with 3FGL\,J0708.9+2239. Bottom: The same spectrum, normalised to highlight features (if any). If present, telluric lines are marked with $\oplus$, and features due to contamination from diffuse interstellar bands are marked with DIB. If there are any doublets, these are marked with a d. The absorption line at $\sim$5890\AA\, which is NaI from the Milky Way, is marked as NaIMW. Unidentified lines are marked with a question mark. (Right panel)  The finding chart ( 5'$\times$ 5' ) retrieved from the Digitized Sky Survey highlighting the location of the optical source: WISE\,J070858.28+224135.4 (red circle)} 
 \label{fig:14}
 \end{figure}
  
    \end{onecolumn}
 \newpage
 \begin{onecolumn}
 
 \centering
 \begin{figure}[t]
 \includegraphics[scale=1.0]{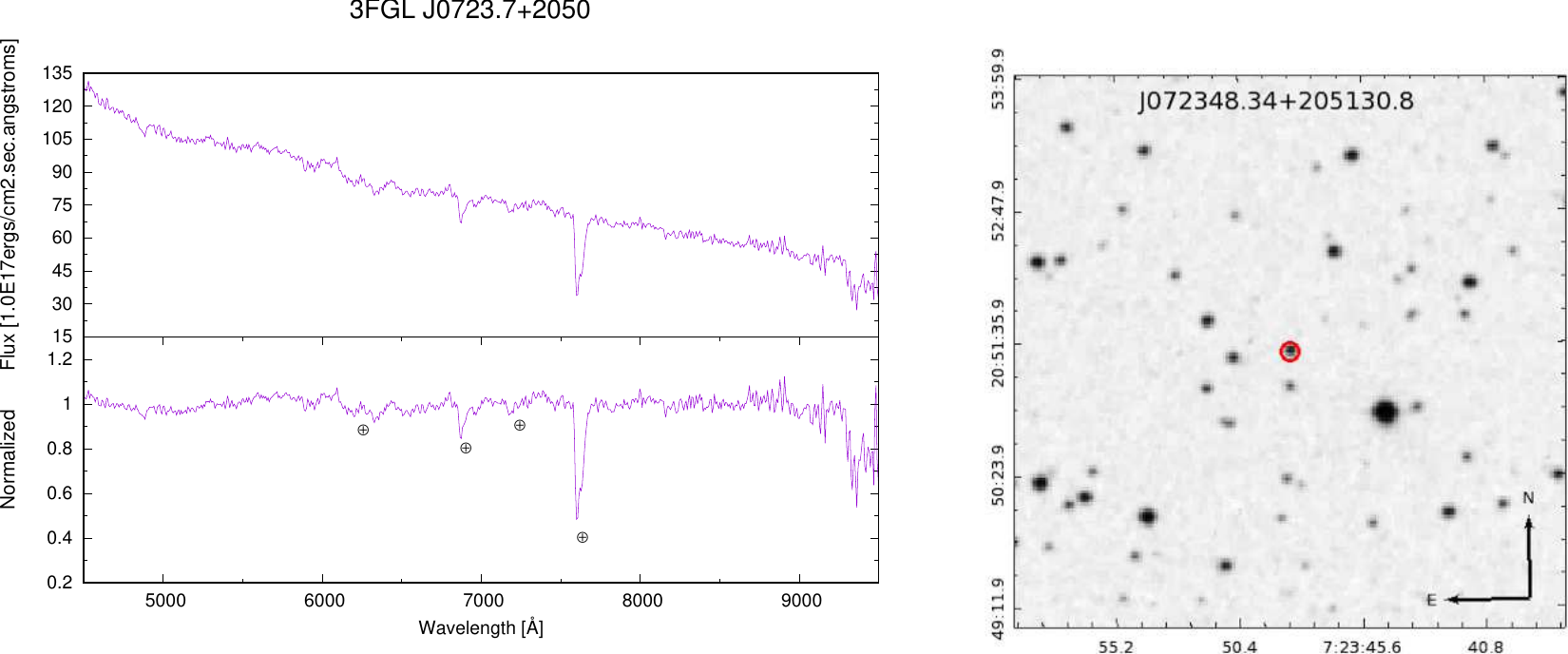}
 \caption{(Left panel) Top: Optical spectrum of WISE\,J072348.34+205130.8 associated with 3FGL\,J0723.7+2050. Bottom: The same spectrum, normalised to highlight features (if any). If present, telluric lines are marked with $\oplus$, and features due to contamination from diffuse interstellar bands are marked with DIB. If there are any doublets, these are marked with a d. The absorption line at $\sim$5890\AA\, which is NaI from the Milky Way, is marked as NaIMW. Unidentified lines are marked with a question mark. (Right panel)  The finding chart ( 5'$\times$ 5' ) retrieved from the Digitized Sky Survey highlighting the location of the optical source: WISE\,J072348.34+205130.8 (red circle)} 
 \label{fig:15}
 \end{figure}
 \centering

 \begin{figure}[b]
 \includegraphics[scale=1.0]{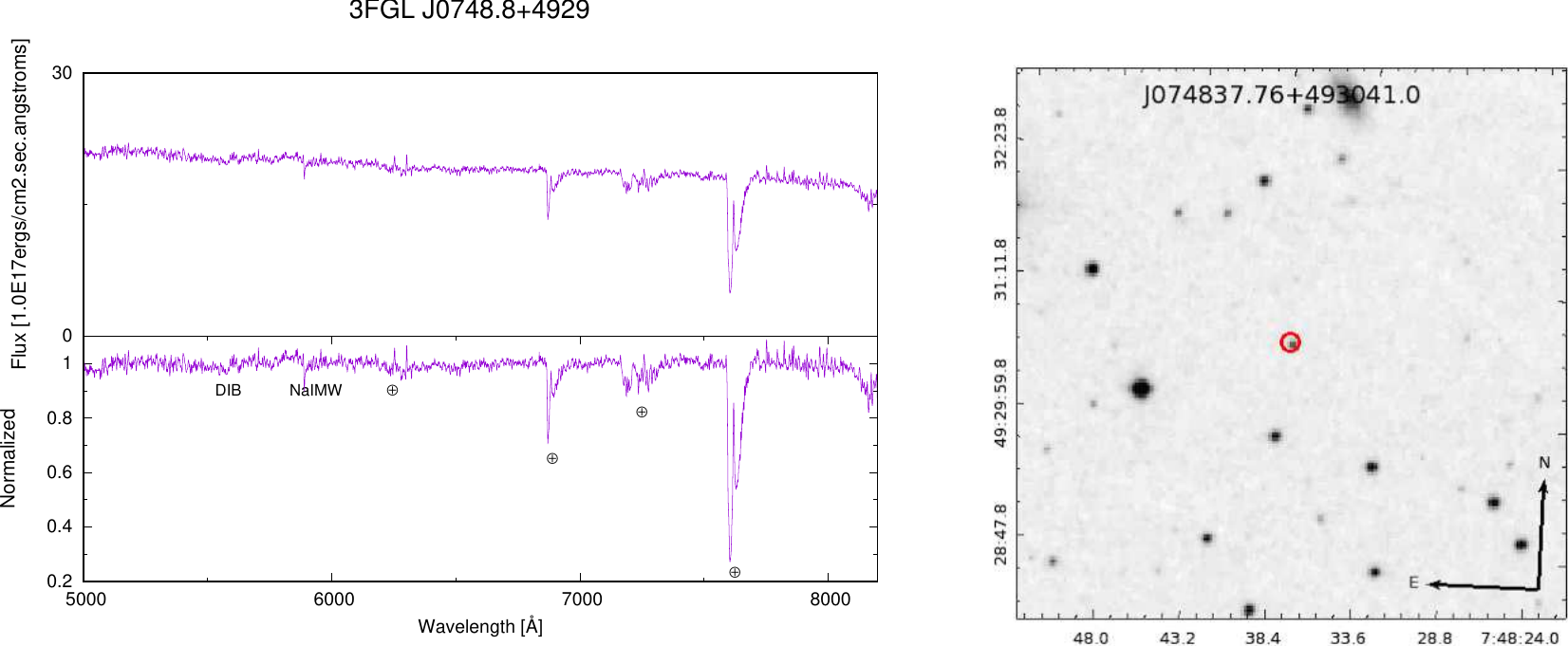}
 \caption{(Left panel) Top: Optical spectrum of WISE\,J074837.76+493041.0 associated with 3FGL\,J0748.8+4929. Bottom: The same spectrum, normalised to highlight features (if any). If present, telluric lines are marked with $\oplus$, and features due to contamination from diffuse interstellar bands are marked with DIB. If there are any doublets, these are marked with a d. The absorption line at $\sim$5890\AA\, which is NaI from the Milky Way, is marked as NaIMW. Unidentified lines are marked with a question mark. (Right panel)  The finding chart ( 5'$\times$ 5' ) retrieved from the Digitized Sky Survey highlighting the location of the optical source: WISE\,J074837.76+493041.0 (red circle)} 
 \label{fig:16}
 \end{figure}
  
    \end{onecolumn}
 \newpage
  \begin{onecolumn}

 \begin{figure}[b]
 \includegraphics[scale=1.0]{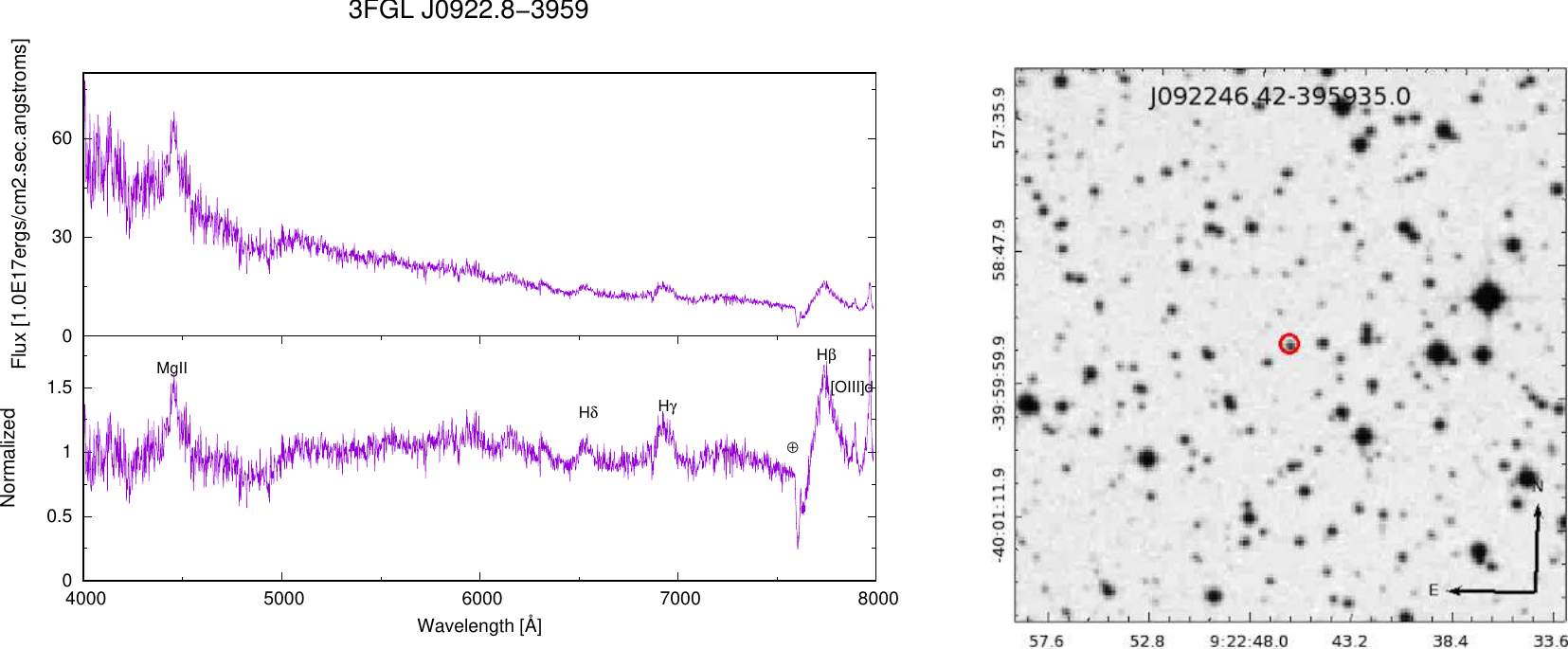}
 \caption{(Left panel) Top: Optical spectrum of WISE\,J092246.42-395935.0 associated with 3FGL\,J0922.8-3959. Bottom: The same spectrum, normalised to highlight features (if any). If present, telluric lines are marked with $\oplus$, and features due to contamination from diffuse interstellar bands are marked with DIB. If there are any doublets, these are marked with a d. The absorption line at $\sim$5890\AA\, which is NaI from the Milky Way, is marked as NaIMW. Unidentified lines are marked with a question mark. (Right panel)  The finding chart ( 5'$\times$ 5' ) retrieved from the Digitized Sky Survey highlighting the location of the optical source: WISE\,J092246.42-395935.0 (red circle)} 
 \label{fig:17}
 \end{figure}

 \centering
 \begin{figure}[t]
 \includegraphics[scale=1.0]{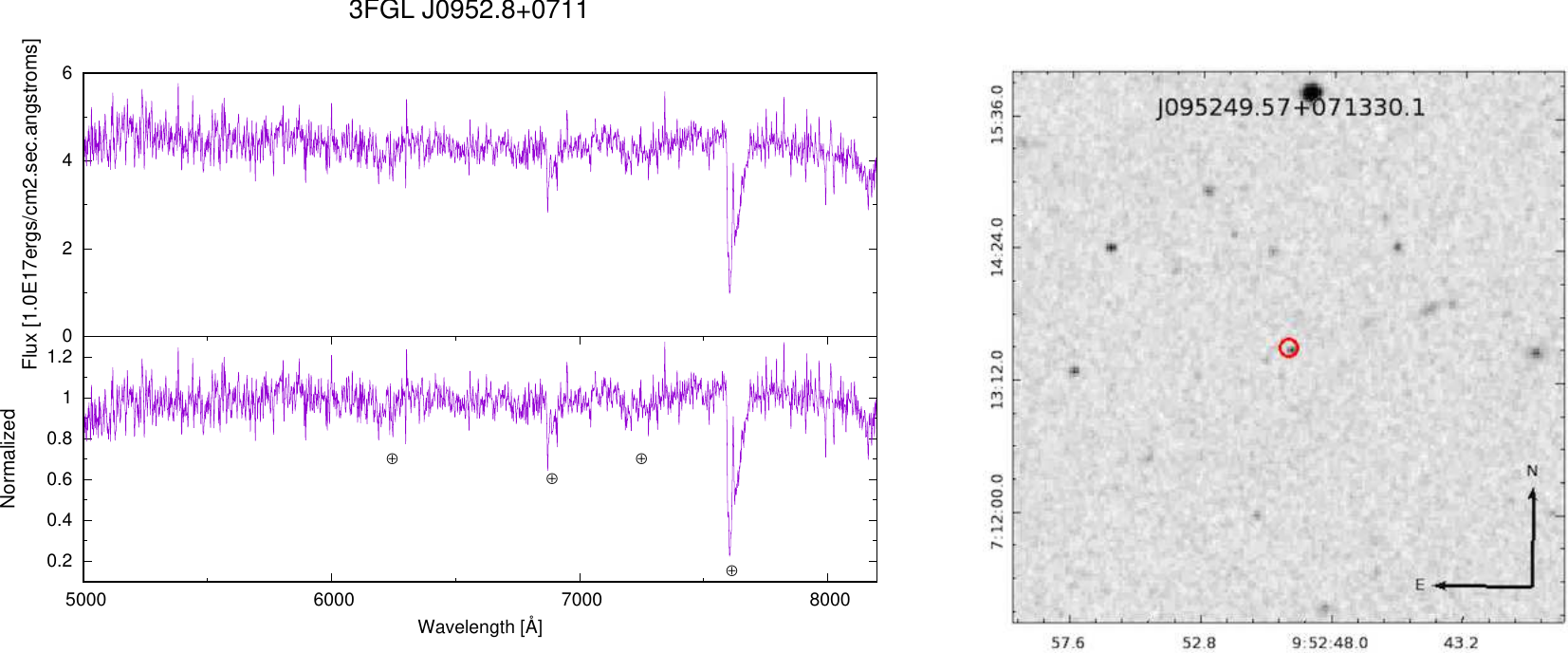}
 \caption{(Left panel) Top: Optical spectrum of WISE\,J095249.57+071330.1 associated with 3FGL\,J0952.8+0711. Bottom: The same spectrum, normalised to highlight features (if any). If present, telluric lines are marked with $\oplus$, and features due to contamination from diffuse interstellar bands are marked with DIB. If there are any doublets, these are marked with a d. The absorption line at $\sim$5890\AA\, which is NaI from the Milky Way, is marked as NaIMW. Unidentified lines are marked with a question mark. (Right panel)  The finding chart ( 5'$\times$ 5' ) retrieved from the Digitized Sky Survey highlighting the location of the optical source: WISE\,J095249.57+071330.1 (red circle)} 
 \label{fig:18}
 \end{figure}
 
     \end{onecolumn}
 \newpage
  \begin{onecolumn}

 \centering
 \begin{figure}[t]
 \includegraphics[scale=1.0]{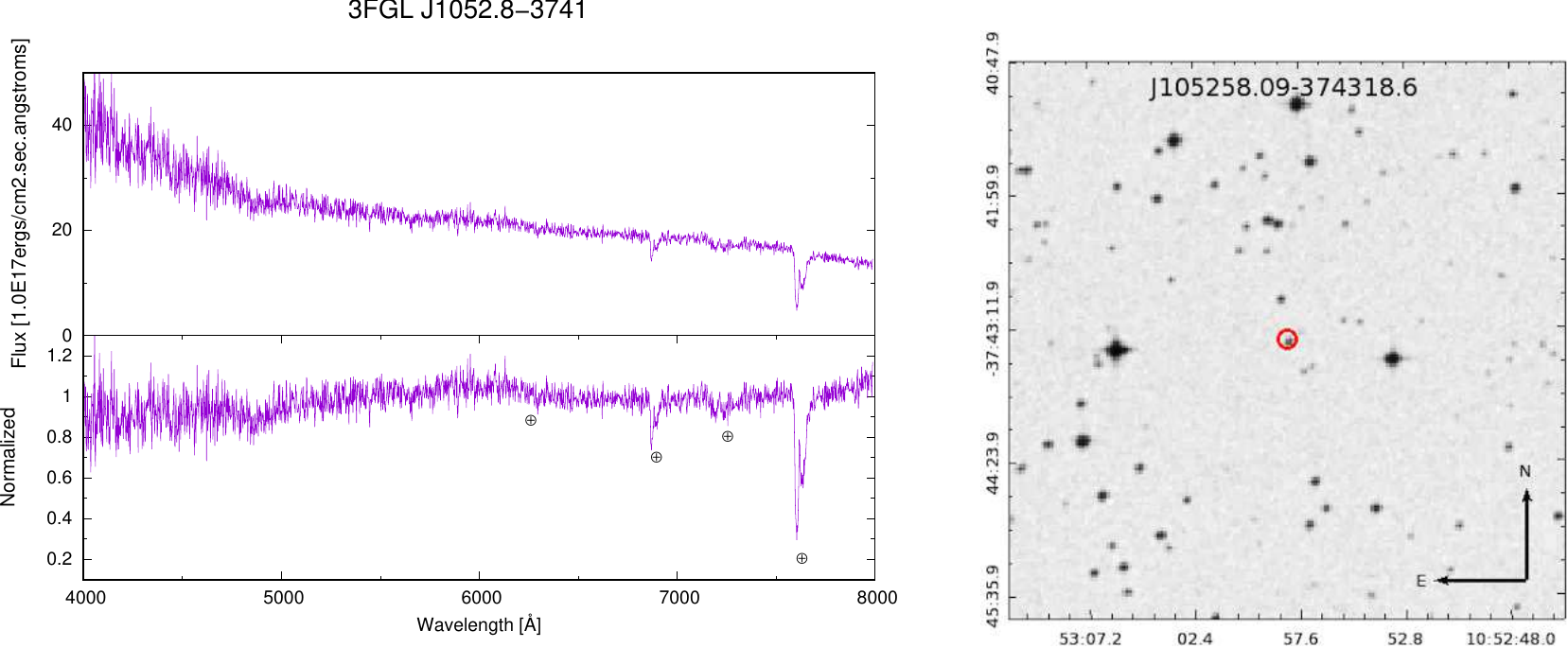}
 \caption{(Left panel) Top: Optical spectrum of WISE\,J105258.09-374318.6 associated with 3FGL\,J1052.8-3741. Bottom: The same spectrum, normalised to highlight features (if any). If present, telluric lines are marked with $\oplus$, and features due to contamination from diffuse interstellar bands are marked with DIB. If there are any doublets, these are marked with a d. The absorption line at $\sim$5890\AA\, which is NaI from the Milky Way, is marked as NaIMW. Unidentified lines are marked with a question mark. (Right panel)  The finding chart ( 5'$\times$ 5' ) retrieved from the Digitized Sky Survey highlighting the location of the optical source: WISE\,J105258.09-374318.6 (red circle)} 
 \label{fig:19}
 \end{figure}

 \centering
 \begin{figure}[t]
 \includegraphics[scale=1.0]{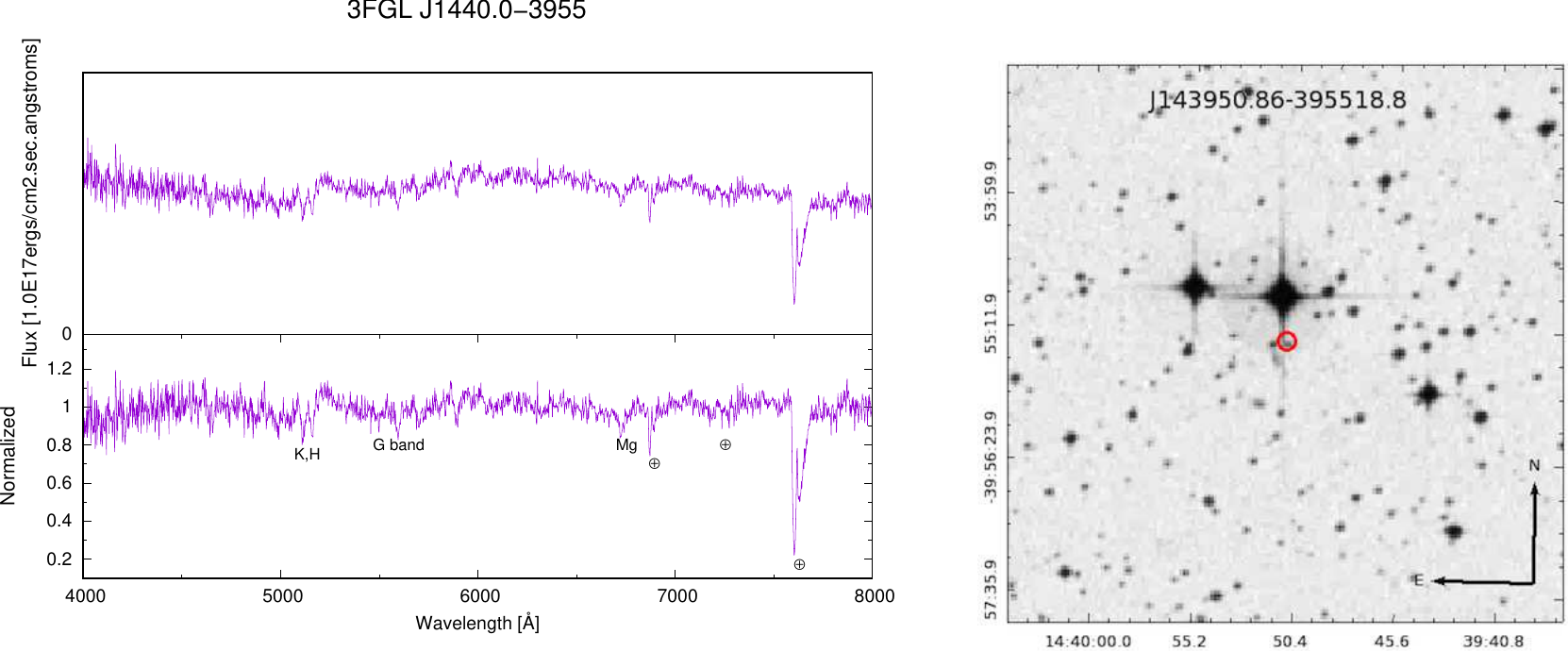}
 \caption{(Left panel) Top: Optical spectrum of WISE\,J143950.86-395518.8 associated with 3FGL\,J1440.0-3955. Bottom: The same spectrum, normalised to highlight features (if any). If present, telluric lines are marked with $\oplus$, and features due to contamination from diffuse interstellar bands are marked with DIB. If there are any doublets, these are marked with a d. The absorption line at $\sim$5890\AA\, which is NaI from the Milky Way, is marked as NaIMW. Unidentified lines are marked with a question mark. (Right panel)  The finding chart ( 5'$\times$ 5' ) retrieved from the Digitized Sky Survey highlighting the location of the optical source: WISE\,J143950.86-395518.8 (red circle)} 
 \label{fig:20}
 \end{figure}

      \end{onecolumn}
 \newpage
  \begin{onecolumn}
  
 \centering
 \begin{figure}[b]
 \includegraphics[scale=1.0]{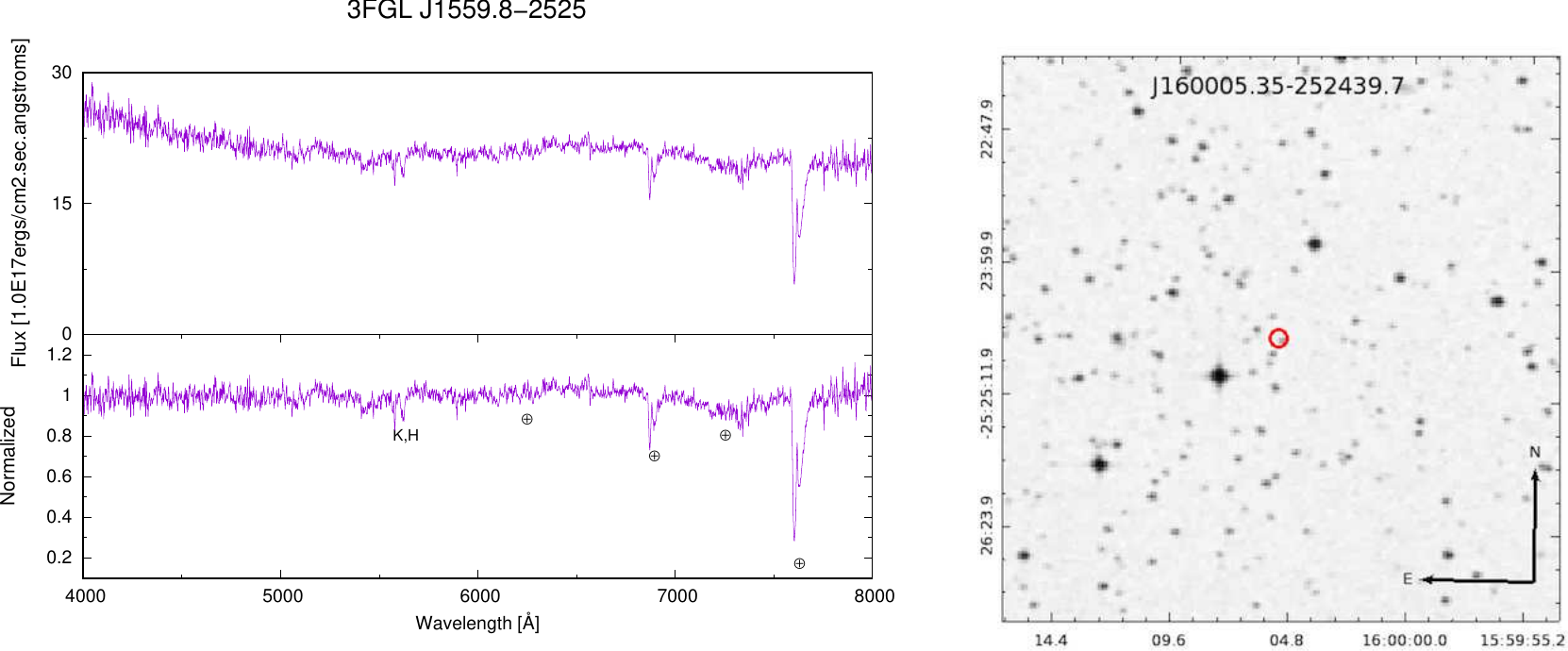}
 \caption{(Left panel) Top: Optical spectrum of WISE\,J160005.35-252439.7 associated with 3FGL\,J1559.8-2525. Bottom: The same spectrum, normalised to highlight features (if any). If present, telluric lines are marked with $\oplus$, and features due to contamination from diffuse interstellar bands are marked with DIB. If there are any doublets, these are marked with a d. The absorption line at $\sim$5890\AA\, which is NaI from the Milky Way, is marked as NaIMW. Unidentified lines are marked with a question mark. (Right panel)  The finding chart ( 5'$\times$ 5' ) retrieved from the Digitized Sky Survey highlighting the location of the optical source: WISE\,J160005.35-252439.7 (red circle)} 
 \label{fig:21}
 \end{figure}

 \centering
 \begin{figure}[t]
 \includegraphics[scale=1.0]{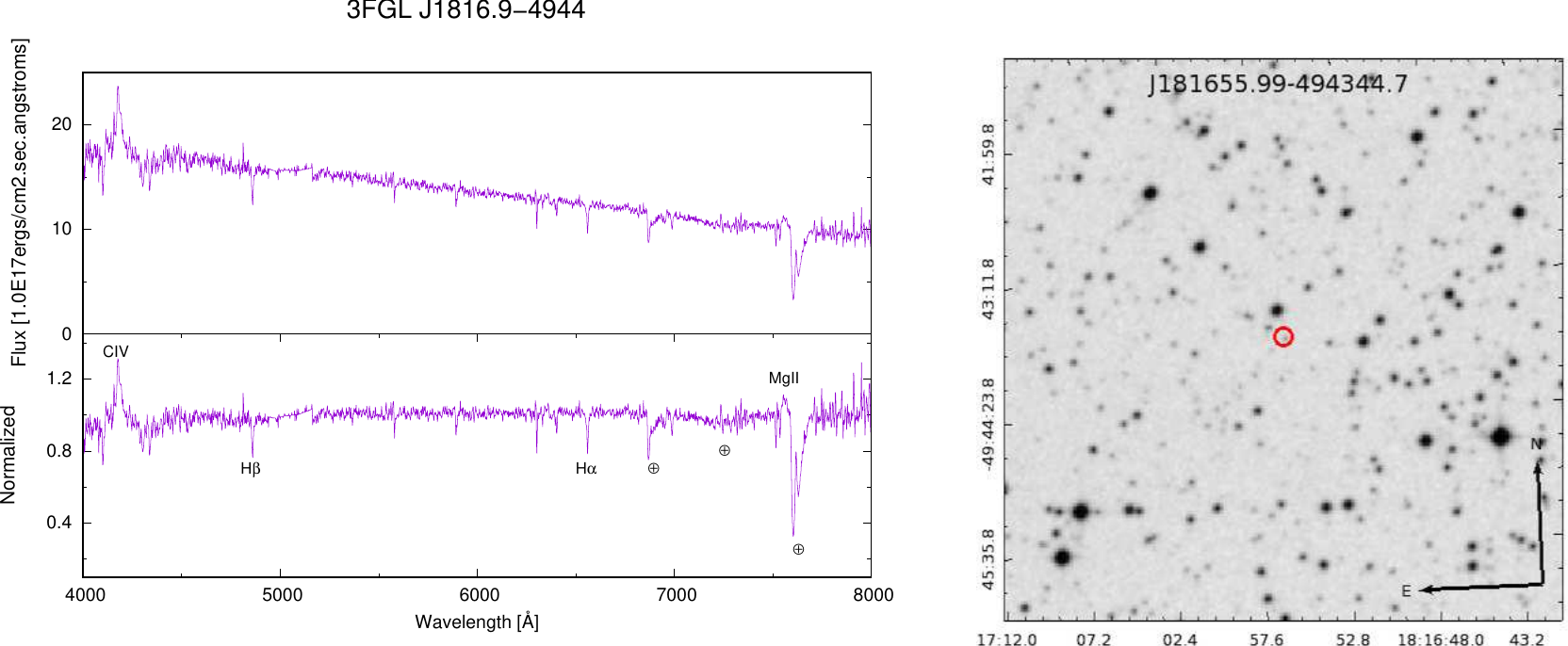}
 \caption{(Left panel) Top: Optical spectrum of WISE\,J181655.99-494344.7 associated with 3FGL\,J1816.9-4944. Bottom: The same spectrum, normalised to highlight features (if any). If present, telluric lines are marked with $\oplus$, and features due to contamination from diffuse interstellar bands are marked with DIB. If there are any doublets, these are marked with a d. The absorption line at $\sim$5890\AA\, which is NaI from the Milky Way, is marked as NaIMW. Unidentified lines are marked with a question mark. (Right panel)  The finding chart ( 5'$\times$ 5' ) retrieved from the Digitized Sky Survey highlighting the location of the optical source: WISE\,J181655.99-494344.7 (red circle). It is worth mentioning there are Balmer lines at $z=0$ due to an interloping star} 
 \label{fig:22}
 \end{figure}
  
      \end{onecolumn}
 \newpage
  \begin{onecolumn}
  
 \centering
 \begin{figure}[b]
 \includegraphics[scale=1.0]{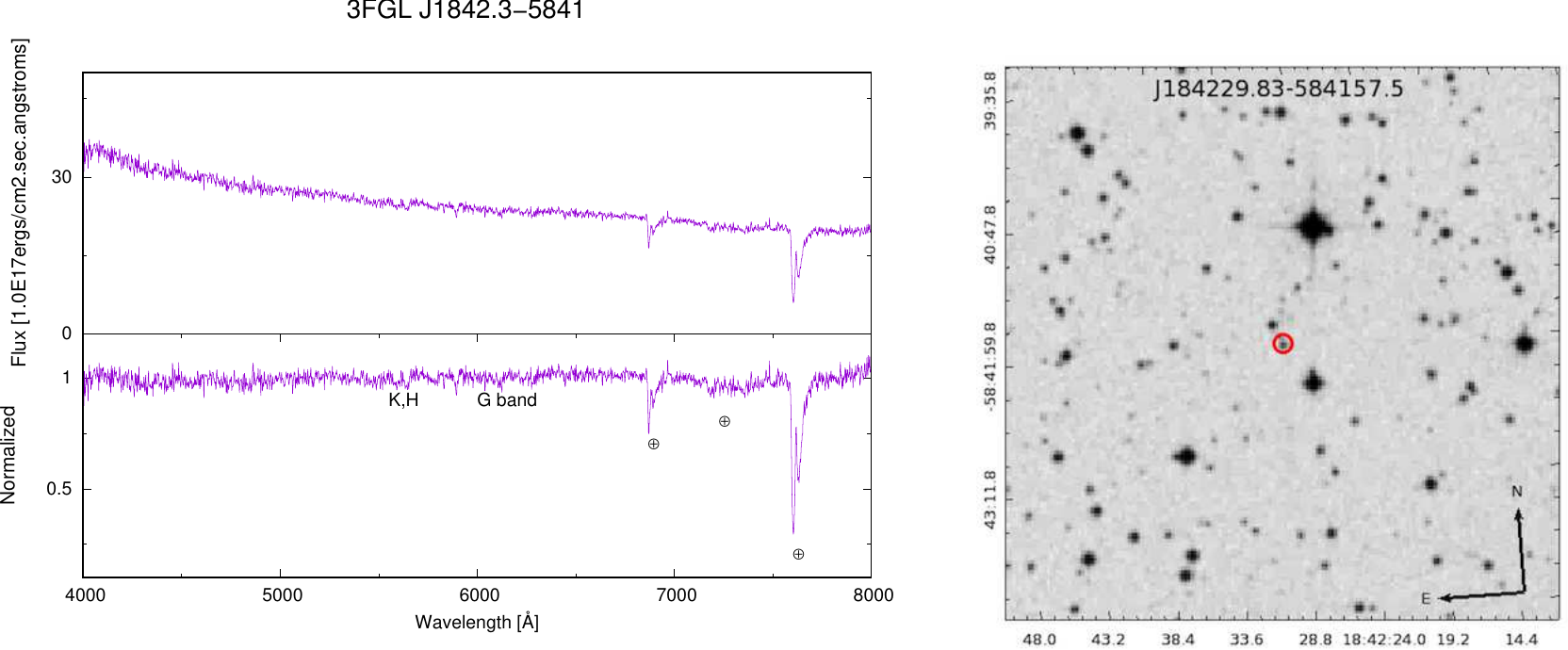}
 \caption{(Left panel) Top: Optical spectrum of WISE\,J184229.83-584157.5 associated with 3FGL\,J1842.3-5841. Bottom: The same spectrum, normalised to highlight features (if any). If present, telluric lines are marked with $\oplus$, and features due to contamination from diffuse interstellar bands are marked with DIB. If there are any doublets, these are marked with a d. The absorption line at $\sim$5890\AA\, which is NaI from the Milky Way, is marked as NaIMW. Unidentified lines are marked with a question mark. (Right panel)  The finding chart ( 5'$\times$ 5' ) retrieved from the Digitized Sky Survey highlighting the location of the optical source: WISE\,J184229.83-584157.5 (red circle)} 
 \label{fig:23}
 \end{figure}

 \centering
 \begin{figure}[t]
 \includegraphics[scale=1.0]{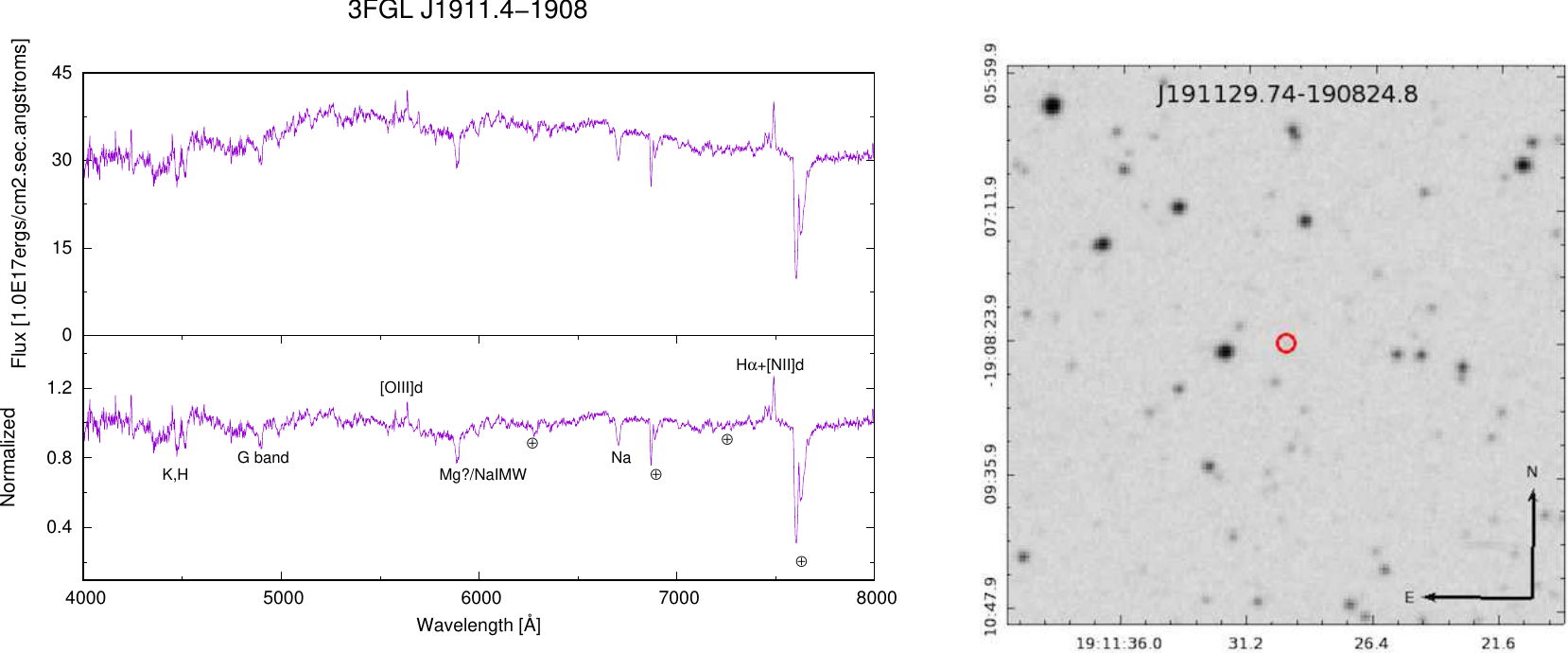}
 \caption{(Left panel) Top: Optical spectrum of WISE\,J191129.74-190824.8 associated with 3FGL\,J1911.4-1908. Bottom: The same spectrum, normalised to highlight features (if any). If present, telluric lines are marked with $\oplus$, and features due to contamination from diffuse interstellar bands are marked with DIB. If there are any doublets, these are marked with a d. The absorption line at $\sim$5890\AA\, which is NaI from the Milky Way, is marked as NaIMW. Unidentified lines are marked with a question mark. (Right panel)  The finding chart ( 5'$\times$ 5' ) retrieved from the Digitized Sky Survey highlighting the location of the optical source: WISE\,J191129.74-190824.8 (red circle)} 
 \label{fig:24}
 \end{figure}
  
      \end{onecolumn}
 \newpage
  \begin{onecolumn}
  
 \centering
 \begin{figure}[b]
 \includegraphics[scale=1.0]{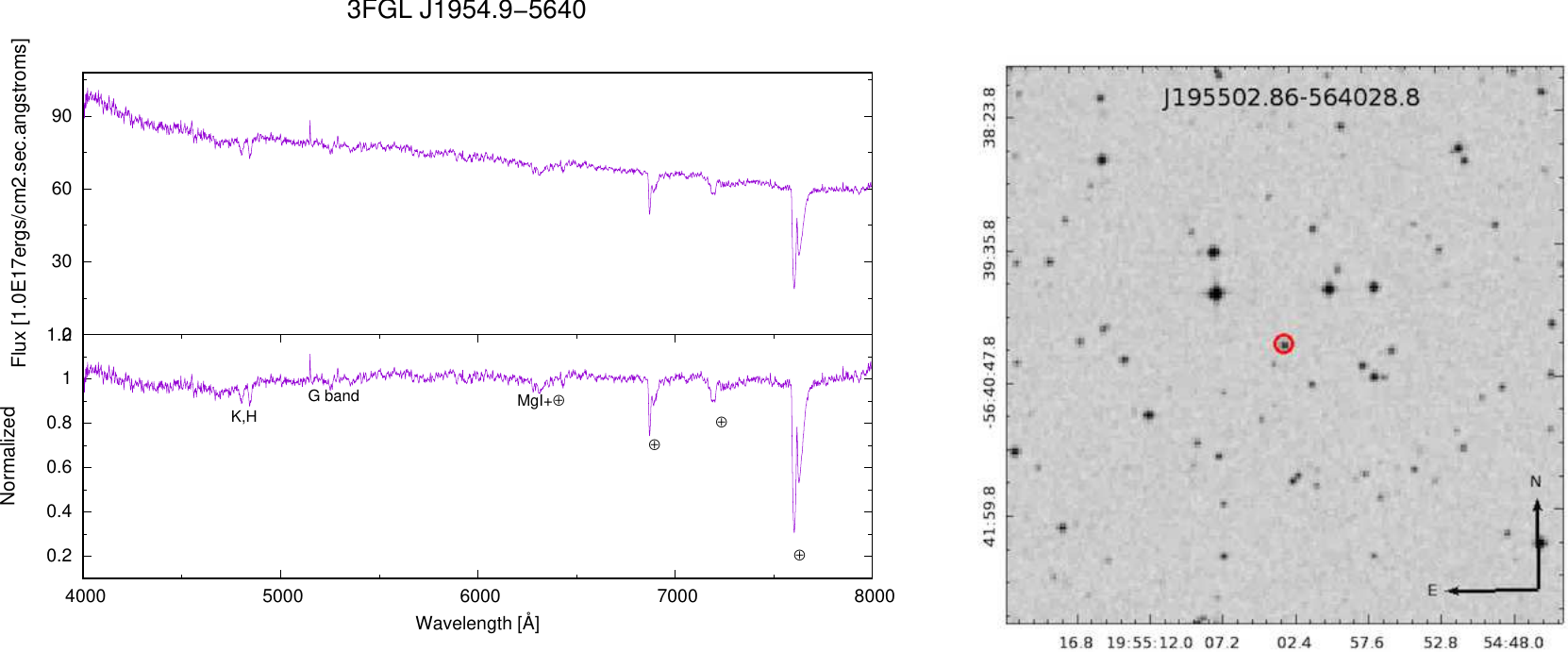}
 \caption{(Left panel) Top: Optical spectrum of WISE\,J195502.86-564028.8 associated with 3FGL\,J1954.9-5640. Bottom: The same spectrum, normalised to highlight features (if any). If present, telluric lines are marked with $\oplus$, and features due to contamination from diffuse interstellar bands are marked with DIB. If there are any doublets, these are marked with a d. The absorption line at $\sim$5890\AA\, which is NaI from the Milky Way, is marked as NaIMW. Unidentified lines are marked with a question mark. (Right panel)  The finding chart ( 5'$\times$ 5' ) retrieved from the Digitized Sky Survey highlighting the location of the optical source: WISE\,J195502.86-564028.8 (red circle)} 
 \label{fig:25}
 \end{figure}

 \centering
 \begin{figure}[t]
 \includegraphics[scale=1.0]{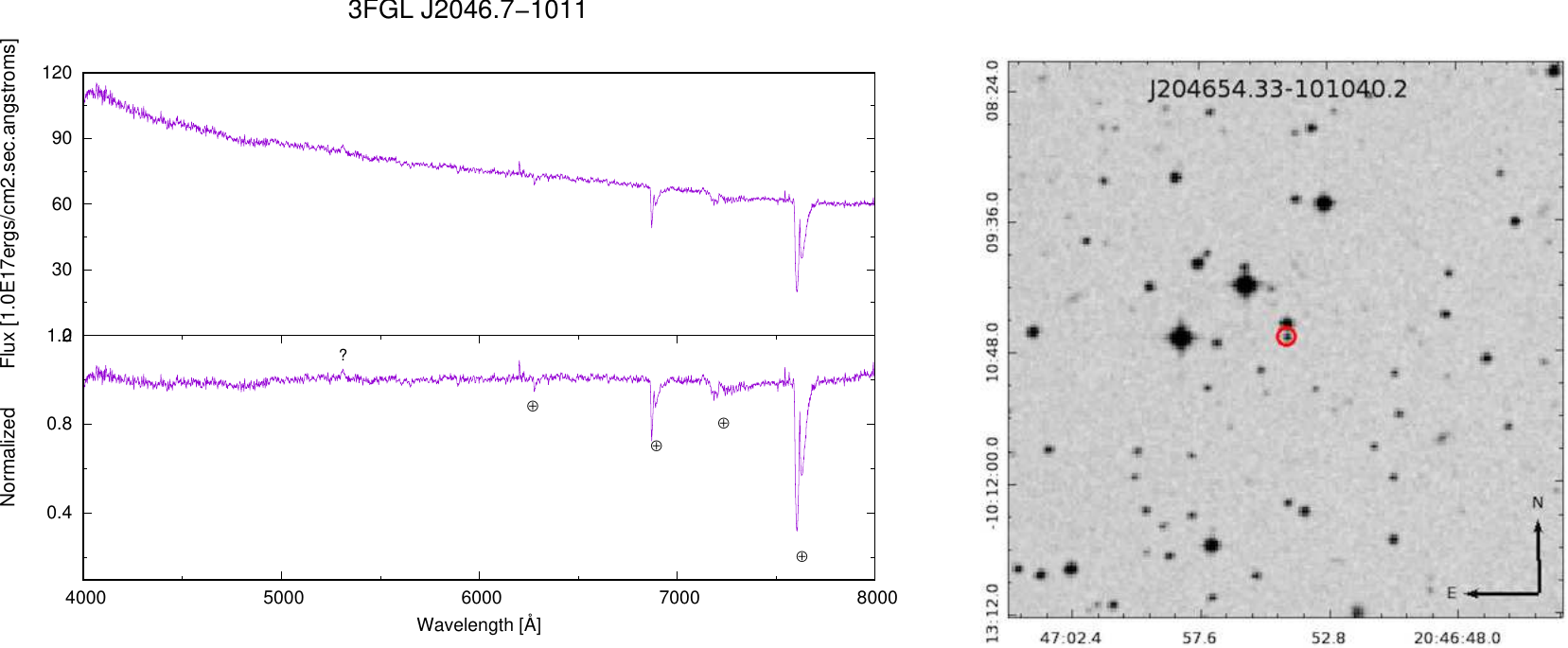}
 \caption{(Left panel) Top: Optical spectrum of WISE\,J204654.33-101040.2 associated with 3FGL\,J2046.7-1011. Bottom: The same spectrum, normalised to highlight features (if any). If present, telluric lines are marked with $\oplus$, and features due to contamination from diffuse interstellar bands are marked with DIB. If there are any doublets, these are marked with a d. The absorption line at $\sim$5890\AA\, which is NaI from the Milky Way, is marked as NaIMW. Unidentified lines are marked with a question mark. (Right panel)  The finding chart ( 5'$\times$ 5' ) retrieved from the Digitized Sky Survey highlighting the location of the optical source: WISE\,J204654.33-101040.2 (red circle)} 
 \label{fig:26}
 \end{figure}
  
      \end{onecolumn}
 \newpage
  \begin{onecolumn}
  
 \centering
 \begin{figure}[b]
 \includegraphics[scale=1.0]{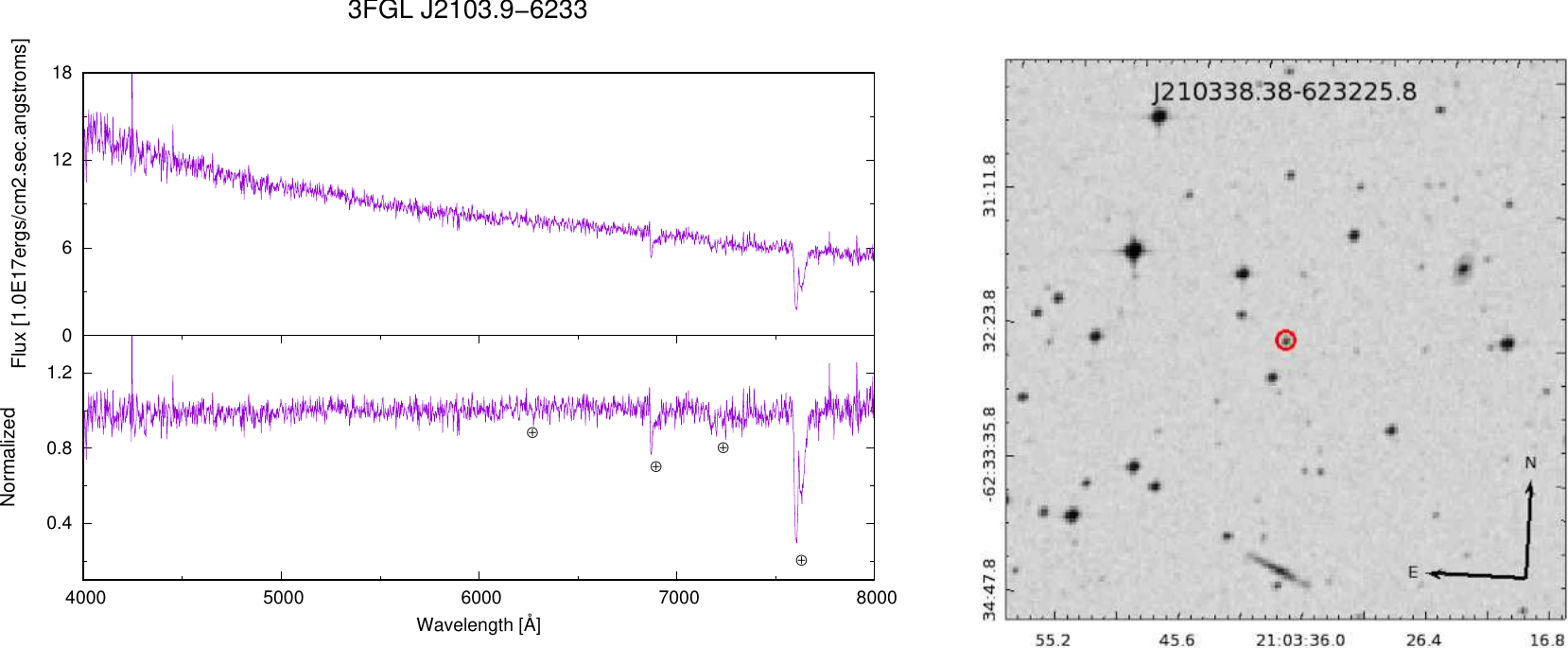}
 \caption{(Left panel) Top: Optical spectrum of WISE\,J210338.38-623225.8 associated with 3FGL\,J2103.9-6233. Bottom: The same spectrum, normalised to highlight features (if any). If present, telluric lines are marked with $\oplus$, and features due to contamination from diffuse interstellar bands are marked with DIB. If there are any doublets, these are marked with a d. The absorption line at $\sim$5890\AA\, which is NaI from the Milky Way, is marked as NaIMW. Unidentified lines are marked with a question mark. (Right panel)  The finding chart ( 5'$\times$ 5' ) retrieved from the Digitized Sky Survey highlighting the location of the optical source: WISE\,J210338.38-623225.8 (red circle)} 
 \label{fig:27}
 \end{figure}

 \centering
 \begin{figure}[t]
 \includegraphics[scale=1.0]{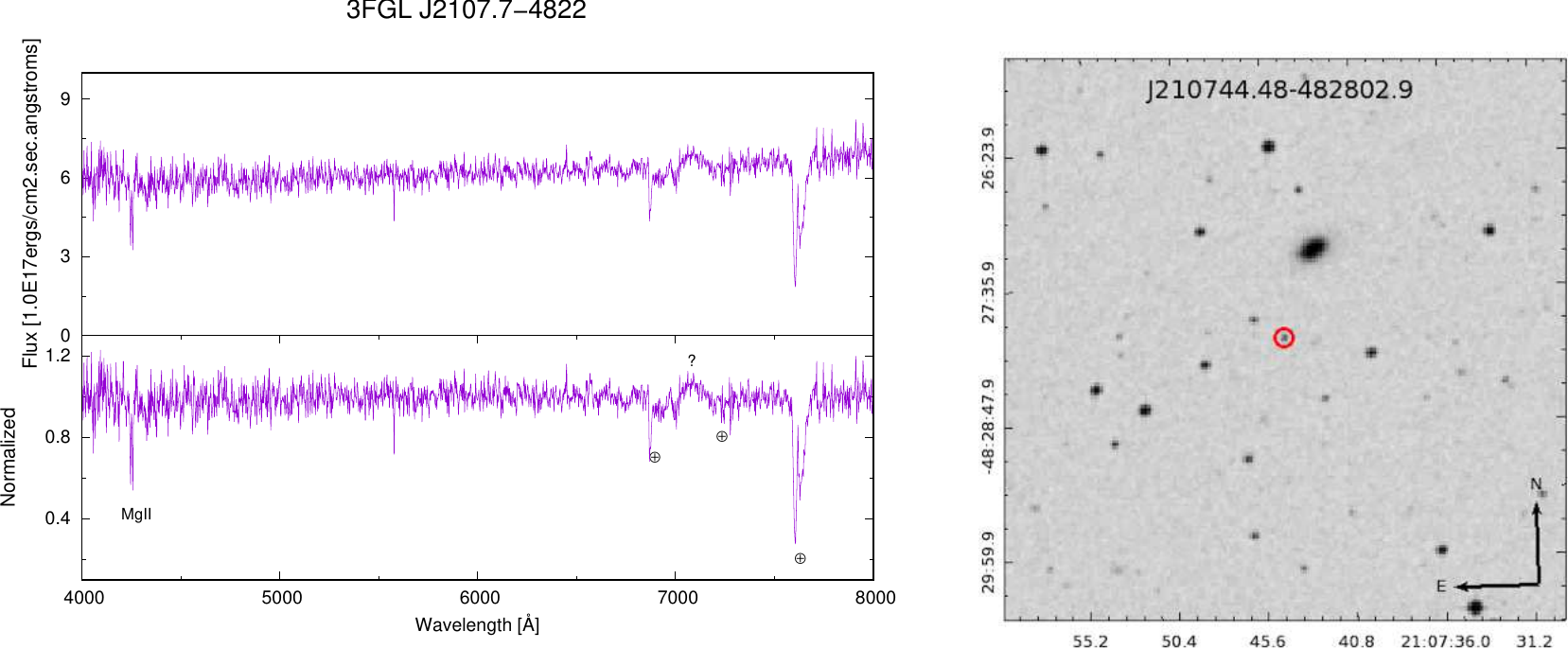}
 \caption{(Left panel) Top: Optical spectrum of WISE\,J210744.48-482802.9 associated with 3FGL\,J2107.7-4822. Bottom: The same spectrum, normalised to highlight features (if any). If present, telluric lines are marked with $\oplus$, and features due to contamination from diffuse interstellar bands are marked with DIB. If there are any doublets, these are marked with a d. The absorption line at $\sim$5890\AA\, which is NaI from the Milky Way, is marked as NaIMW. Unidentified lines are marked with a question mark. (Right panel)  The finding chart ( 5'$\times$ 5' ) retrieved from the Digitized Sky Survey highlighting the location of the optical source: WISE\,J210744.48-482802.9 (red circle)} 
 \label{fig:28}
 \end{figure}
  
      \end{onecolumn}
 \newpage
  \begin{onecolumn}
  
 \centering
 \begin{figure}[b]
 \includegraphics[scale=1.0]{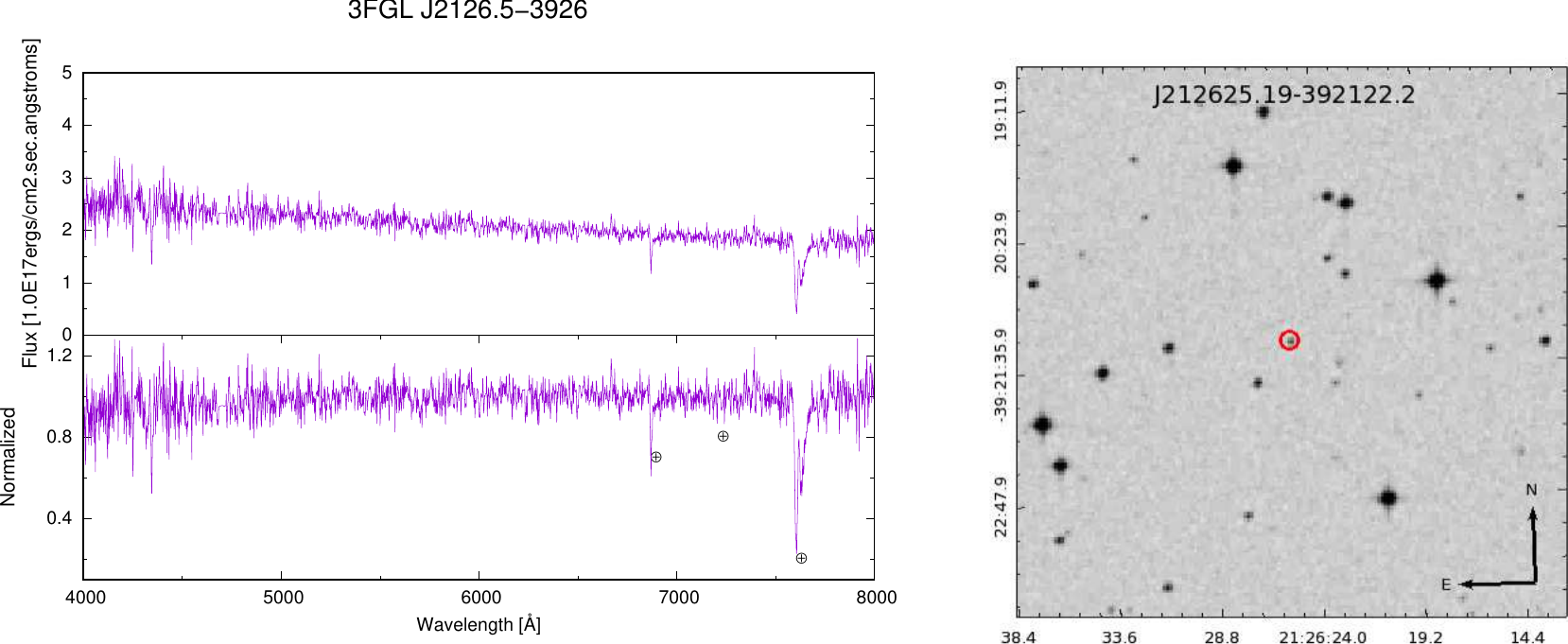}
 \caption{(Left panel) Top: Optical spectrum of WISE\,J212625.19-392122.2 associated with 3FGL\,J2126.5-3926. Bottom: The same spectrum, normalised to highlight features (if any). If present, telluric lines are marked with $\oplus$, and features due to contamination from diffuse interstellar bands are marked with DIB. If there are any doublets, these are marked with a d. The absorption line at $\sim$5890\AA\, which is NaI from the Milky Way, is marked as NaIMW. Unidentified lines are marked with a question mark. (Right panel)  The finding chart ( 5'$\times$ 5' ) retrieved from the Digitized Sky Survey highlighting the location of the optical source: WISE\,J212625.19-392122.2 (red circle)} 
 \label{fig:29}
 \end{figure}

 \centering
 \begin{figure}[t]
 \includegraphics[scale=1.0]{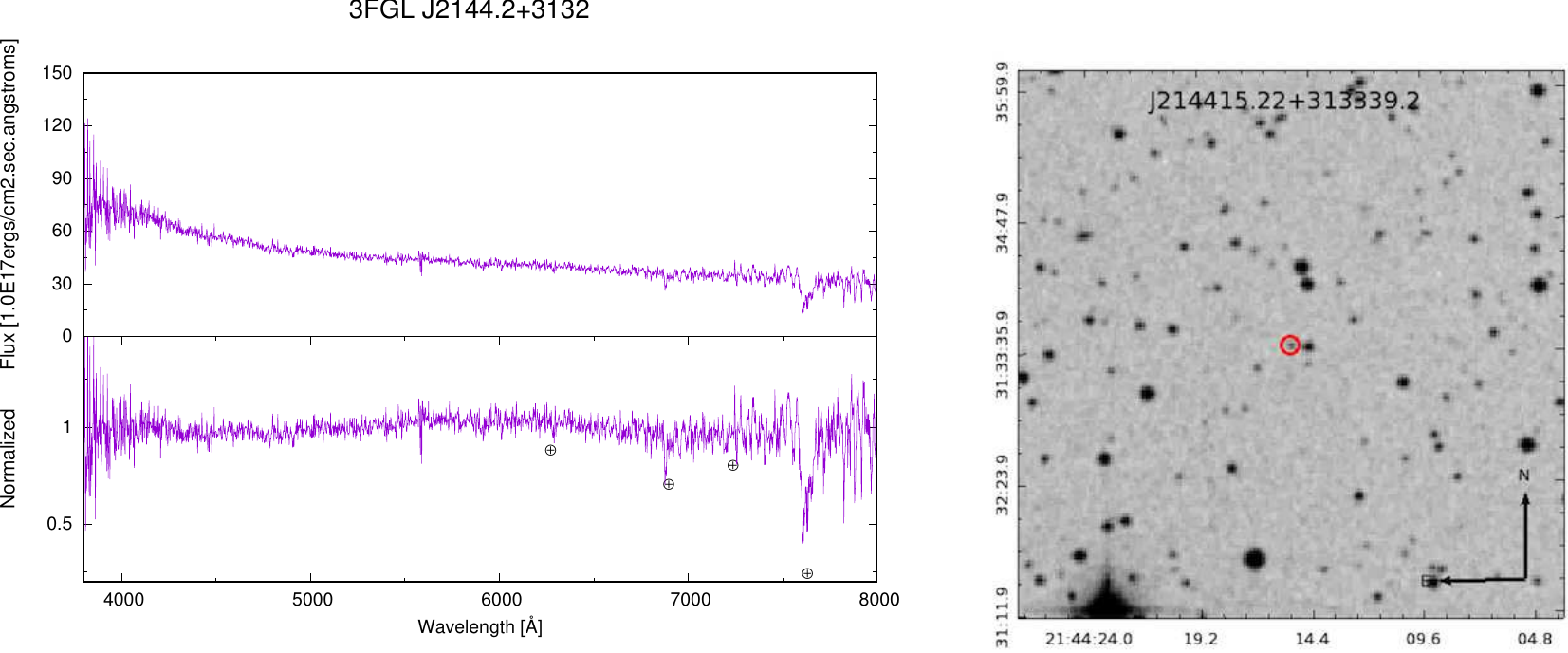}
 \caption{(Left panel) Top: Optical spectrum of WISE\,J214415.22+313339.2 associated with 3FGL\,J2144.2+3132. Bottom: The same spectrum, normalised to highlight features (if any). If present, telluric lines are marked with $\oplus$, and features due to contamination from diffuse interstellar bands are marked with DIB. If there are any doublets, these are marked with a d. The absorption line at $\sim$5890\AA\, which is NaI from the Milky Way, is marked as NaIMW. Unidentified lines are marked with a question mark. (Right panel)  The finding chart ( 5'$\times$ 5' ) retrieved from the Digitized Sky Survey highlighting the location of the optical source: WISE\,J214415.22+313339.2 (red circle)} 
 \label{fig:30}
 \end{figure}
  
      \end{onecolumn}
 \newpage
  \begin{onecolumn}
  
 \centering
 \begin{figure}[b]
 \includegraphics[scale=1.0]{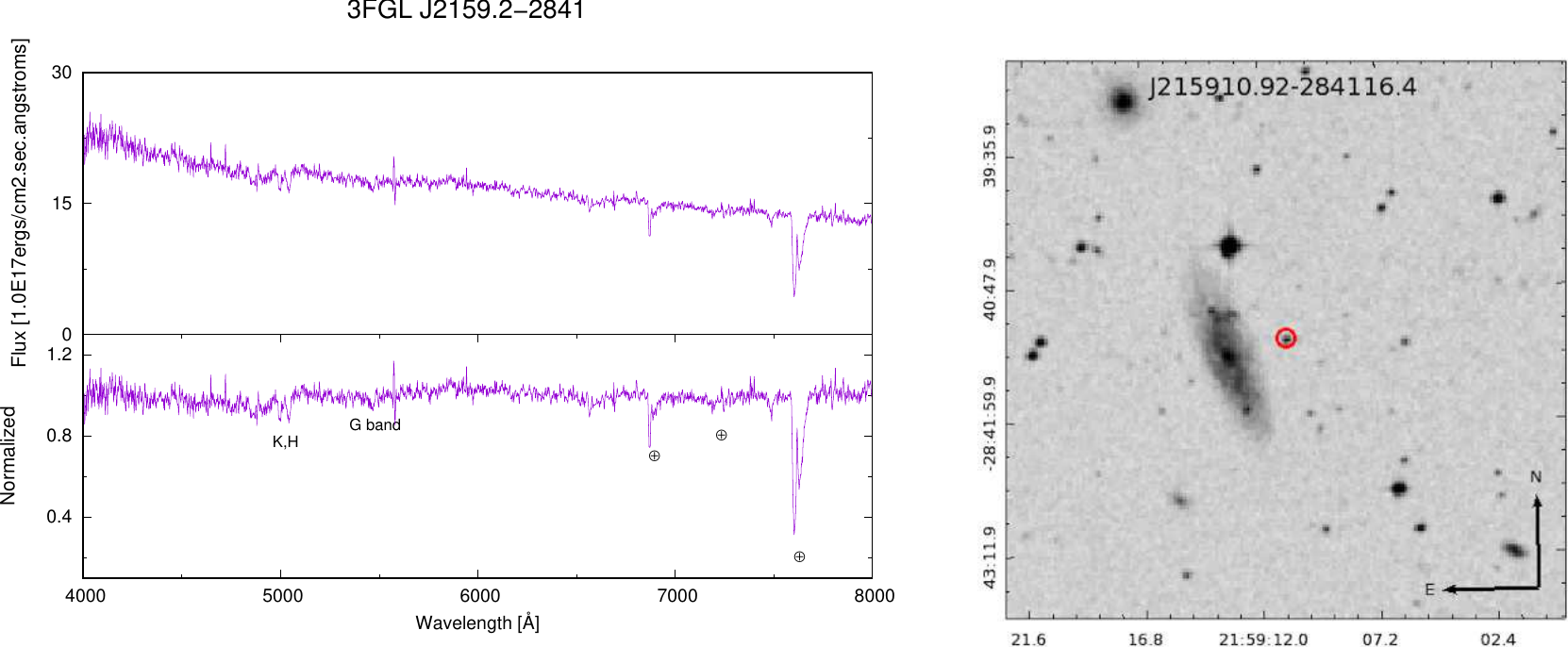}
 \caption{(Left panel) Top: Optical spectrum of WISE\,J215910.92-284116.4 associated with 3FGL\,J2159.2-2841. Bottom: The same spectrum, normalised to highlight features (if any). If present, telluric lines are marked with $\oplus$, and features due to contamination from diffuse interstellar bands are marked with DIB. If there are any doublets, these are marked with a d. The absorption line at $\sim$5890\AA\, which is NaI from the Milky Way, is marked as NaIMW. Unidentified lines are marked with a question mark. (Right panel)  The finding chart ( 5'$\times$ 5' ) retrieved from the Digitized Sky Survey highlighting the location of the optical source: WISE\,J215910.92-284116.4 (red circle)} 
 \label{fig:31}
 \end{figure}

 \centering
 \begin{figure}[t]
 \includegraphics[scale=1.0]{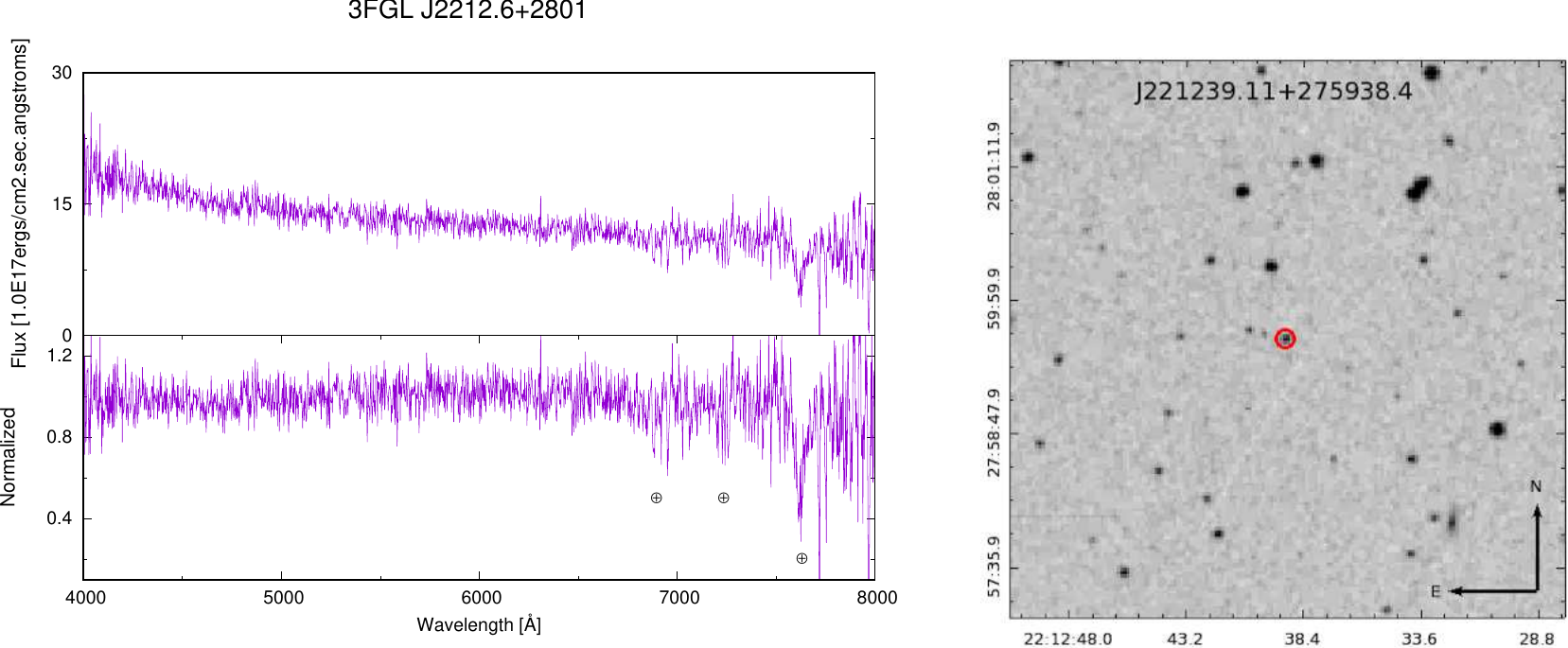}
 \caption{(Left panel) Top: Optical spectrum of WISE\,J221239.1+275938.4 associated with 3FGL\,J2212.6+2801. Bottom: The same spectrum, normalised to highlight features (if any). If present, telluric lines are marked with $\oplus$, and features due to contamination from diffuse interstellar bands are marked with DIB. If there are any doublets, these are marked with a d. The absorption line at $\sim$5890\AA\, which is NaI from the Milky Way, is marked as NaIMW. Unidentified lines are marked with a question mark. (Right panel)  The finding chart ( 5'$\times$ 5' ) retrieved from the Digitized Sky Survey highlighting the location of the optical source: WISE\,J221239.1+275938.4 (red circle)} 
 \label{fig:32}
 \end{figure}
  
      \end{onecolumn}
 \newpage
  \begin{onecolumn}
  
 \centering
 \begin{figure}[b]
 \includegraphics[scale=1.0]{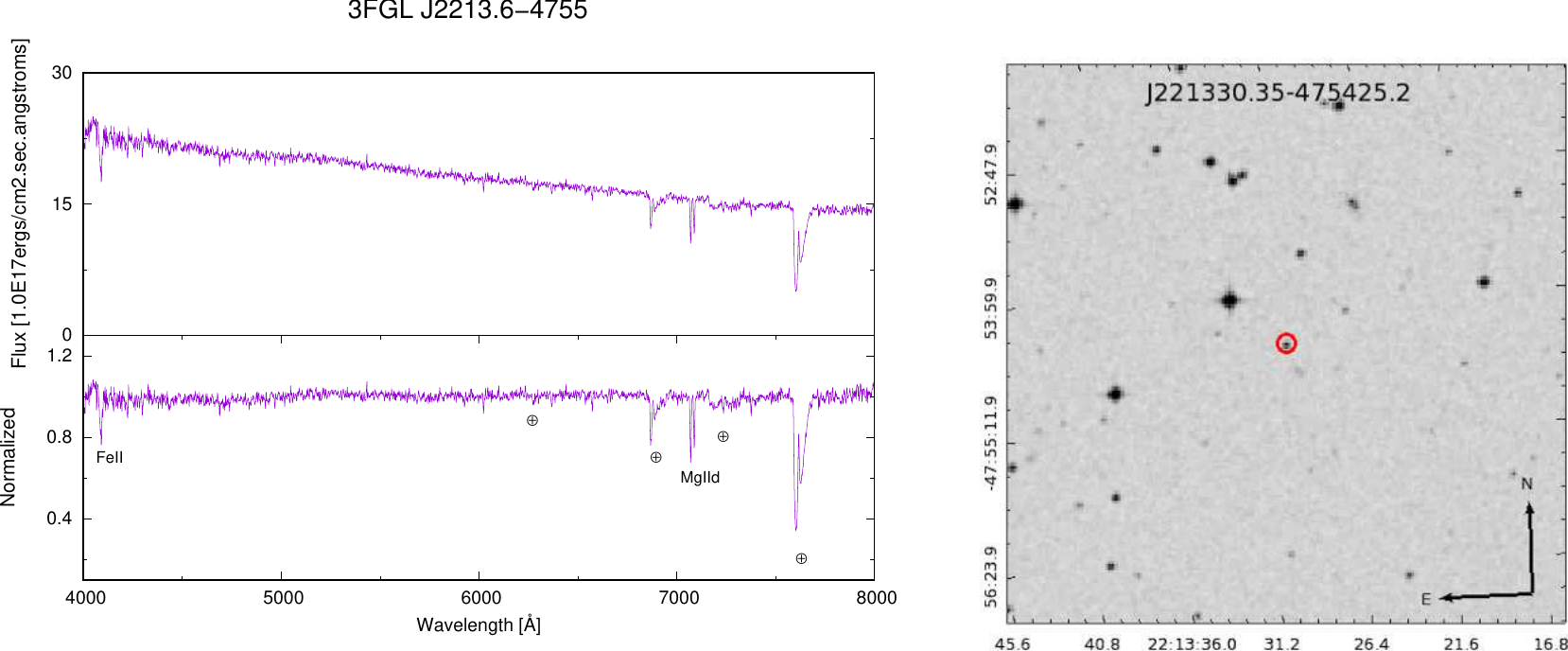}
 \caption{(Left panel) Top: Optical spectrum of WISE\,J221330.35-475425.2 associated with 3FGL\,J2213.6-4755. Bottom: The same spectrum, normalised to highlight features (if any). If present, telluric lines are marked with $\oplus$, and features due to contamination from diffuse interstellar bands are marked with DIB. If there are any doublets, these are marked with a d. The absorption line at $\sim$5890\AA\, which is NaI from the Milky Way, is marked as NaIMW. Unidentified lines are marked with a question mark. (Right panel)  The finding chart ( 5'$\times$ 5' ) retrieved from the Digitized Sky Survey highlighting the location of the optical source: WISE\,J221330.35-475425.2 (red circle)} 
 \label{fig:33}
 \end{figure}

 \centering
 \begin{figure}[t]
 \includegraphics[scale=1.0]{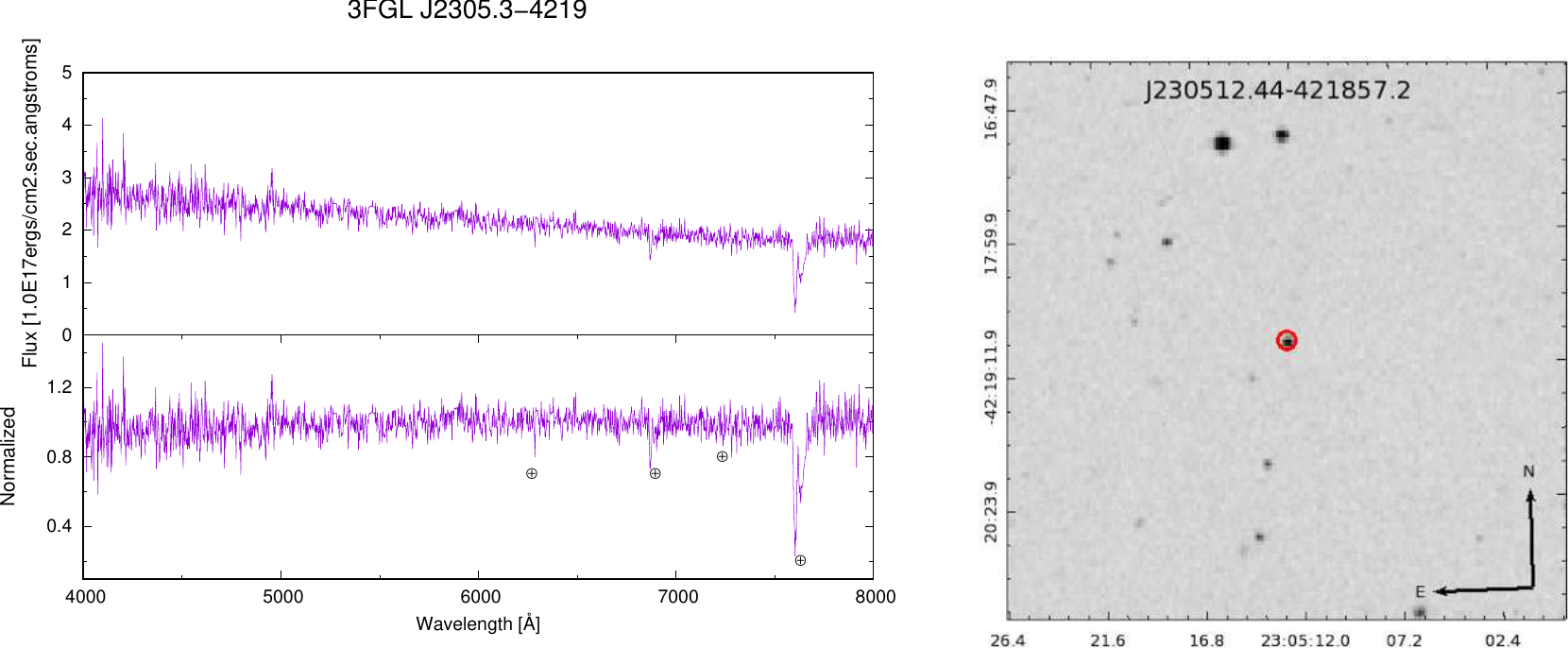}
 \caption{(Left panel) Top: Optical spectrum of WISE\,J230512.44-421857.2 associated with 3FGL\,2305.3-4219. Bottom: The same spectrum, normalised to highlight features (if any). If present, telluric lines are marked with $\oplus$, and features due to contamination from diffuse interstellar bands are marked with DIB. If there are any doublets, these are marked with a d. The absorption line at $\sim$5890\AA\, which is NaI from the Milky Way, is marked as NaIMW. Unidentified lines are marked with a question mark. (Right panel)  The finding chart ( 5'$\times$ 5' ) retrieved from the Digitized Sky Survey highlighting the location of the optical source: WISE\,J230512.44-421857.2 (red circle)} 
 \label{fig:34}
 \end{figure}
  
      \end{onecolumn}
 \newpage
  \begin{onecolumn}
  
 \centering
 \begin{figure}[b]
 \includegraphics[scale=1.0]{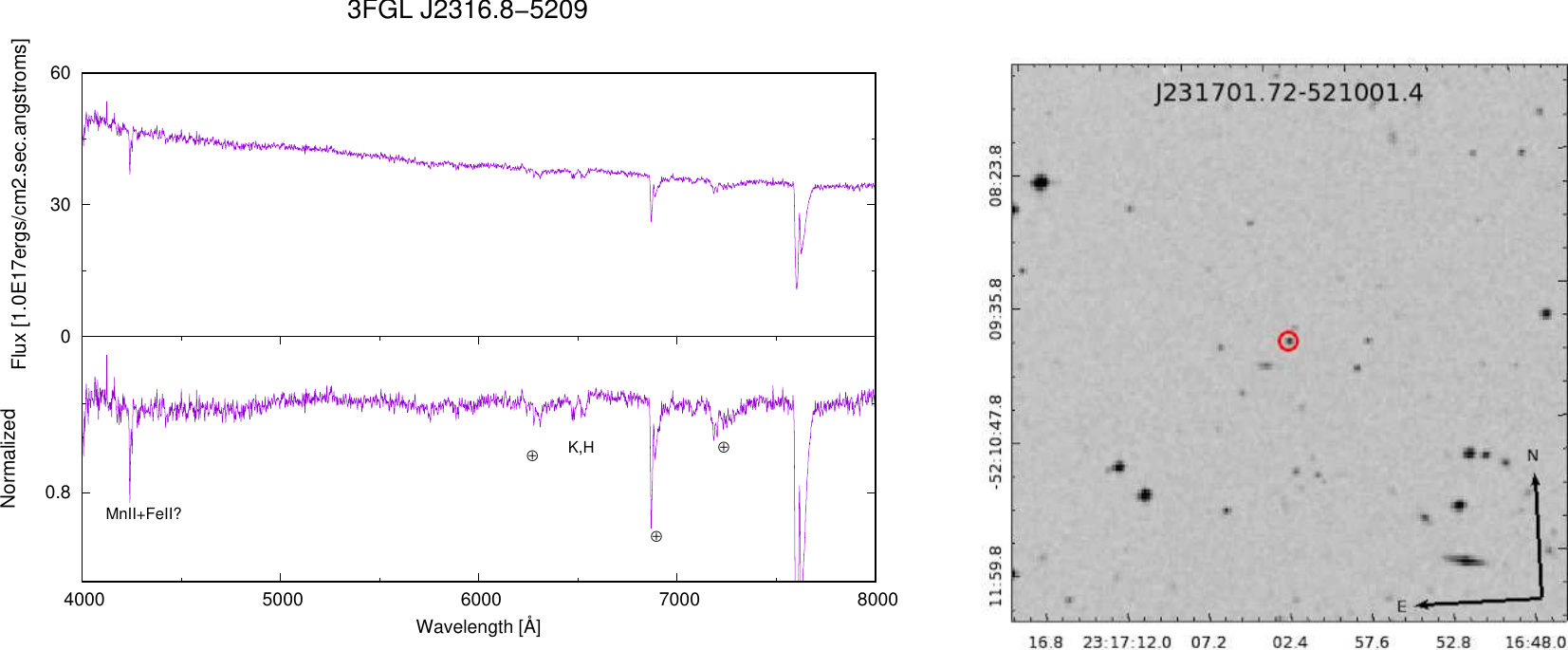}
 \caption{(Left panel) Top: Optical spectrum of WISE\,J231701.72-521001.4 associated with 3FGL\,J2316.8-5209. Bottom: The same spectrum, normalised to highlight features (if any). If present, telluric lines are marked with $\oplus$, and features due to contamination from diffuse interstellar bands are marked with DIB. If there are any doublets, these are marked with a d. The absorption line at $\sim$5890\AA\, which is NaI from the Milky Way, is marked as NaIMW. Unidentified lines are marked with a question mark. (Right panel)  The finding chart ( 5'$\times$ 5' ) retrieved from the Digitized Sky Survey highlighting the location of the optical source: WISE\,J231701.72-521001.4 (red circle)} 
 \label{fig:35}
 \end{figure}

 \centering
 \begin{figure}
\includegraphics[scale=1.0]{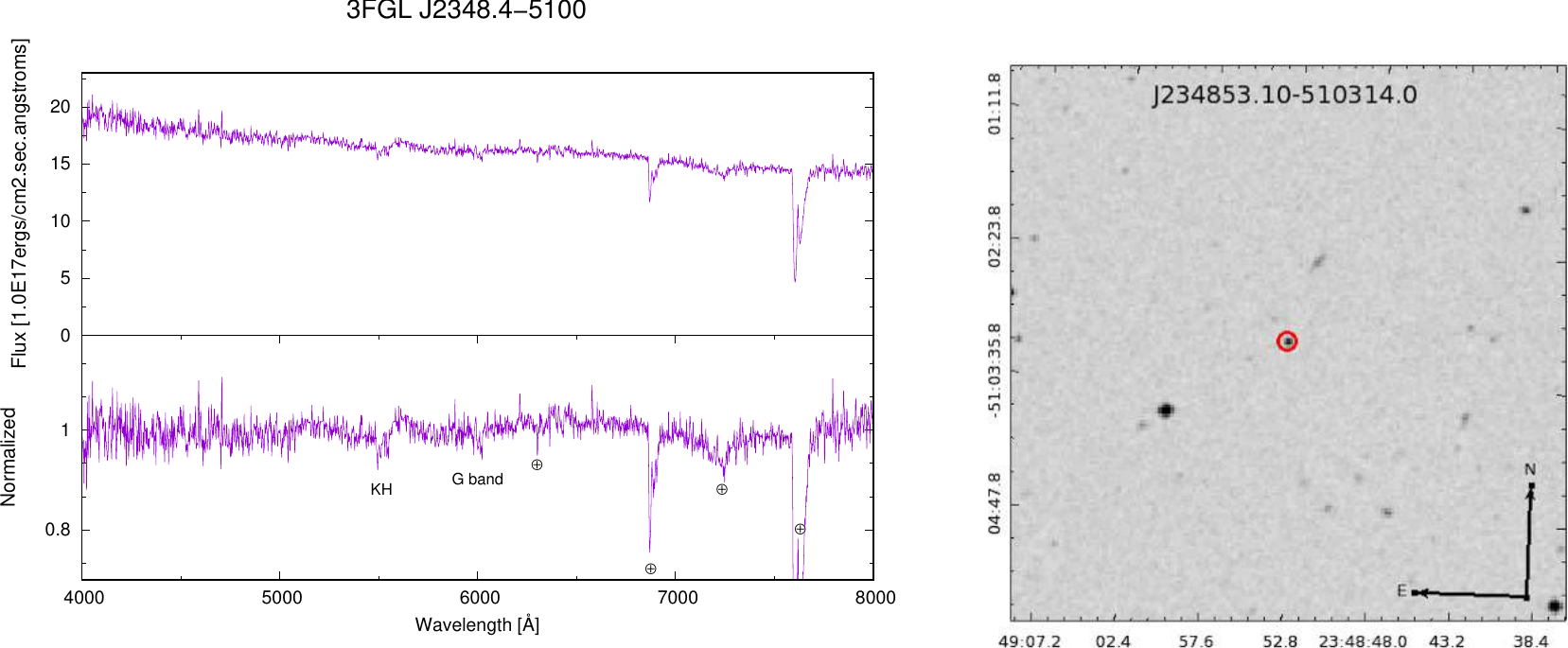}
 \caption{(Left panel) Top: Optical spectrum of WISE\,J234853.10-510314.0 associated with 3FGL\,J2348.4-5100. Bottom: The same spectrum, normalised to highlight features (if any). If present, telluric lines are marked with $\oplus$, and features due to contamination from diffuse interstellar bands are marked with DIB. If there are any doublets, these are marked with a d. The absorption line at $\sim$5890\AA\, which is NaI from the Milky Way, is marked as NaIMW. Unidentified lines are marked with a question mark. (Right panel)  Unidentified lines are marked with a question mark. The finding chart ( 5'$\times$ 5' ) retrieved from the Digitized Sky Survey highlighting the location of the optical source: WISE\,J234853.10-510314.0 (red circle)} 
 \label{fig:36}
 \end{figure}
 
\end{onecolumn}

%
\acknowledgments
This work is supported by the "Departments of Excellence 2018 - 2022" Grant awarded by the Italian Ministry of Education, University and Research (MIUR) (L. 232/2016). This research has made use of resources provided by the Compagnia di San Paolo for the grant awarded on the BLENV project (S1618\_L1\_MASF\_01) and by the Ministry of Education, Universities and Research for the grant MASF\_FFABR\_17\_01. E. J. M. acknowledges the financial contribution from the Universidad Nacional de La Plata - Ay. Mov. Doctorandos 2018. F. M. acknowledges financial contribution from the agreement ASI-INAF n.2017-14-H.0. A. P. acknowledges financial support from the Consorzio Interuniversitario per la fisica Spaziale (CIFS) under the agreement related to the grant MASF\_CONTR\_FIN\_18\_02. F. R. acknowledges the support from FONDECYT Postdoctorado 3180506 and CONICYT project Basal AFB-170002. E. J. B. acknowledges support from Programa de Apoyo a Proyectos de Investigaci\'on e Innovaci\'on Tecnol\'ogica (IN109217).


\end{document}